\documentclass[letter]{aa} % for the letters 
\usepackage{graphicx}
\usepackage[varg]{txfonts}
\usepackage{upgreek} 
\usepackage{natbib}
\usepackage{mhchem}

\begin{document} 

\title{Doubly substituted isotopologues of HCCCN in TMC-1: Detection of
D$^{13}$CCCN, DC$^{13}$CCN, DCC$^{13}$CN, DCCC$^{15}$N, H$^{13}$C$^{13}$CCN, H$^{13}$CC$^{13}$CN, HC$^{13}$C$^{13}$CN, 
HCC$^{13}$C$^{15}$N, and HC$^{13}$CC$^{15}$N\thanks{Based
on observations with the Yebes 40m radio telescope (projects 19A003, 20A014, 20D023, 21A011, 21D005, 22A007, 22B029, and 23A024) and the IRAM 30m radiotelescope. The 40m radiotelescope at Yebes Observatory is operated by the Spanish Geographic Institute (IGN, Ministerio de Transportes y Movilidad Sostenible).
IRAM is supported by INSU/CNRS (France), MPG (Germany), and IGN (Spain).}
%: detection of
%D$^{13}$CCCN, DC$^{13}$CCN, DCC$^{13}$CN, DCCC$^{15}$N, H$^{13}$C$^{13}$CCN, H$^{13}$CC$^{13}$CN, HC$^{13}$C$^{13}$CN, HCC$^{13}$C$^{15}$N, and HC$^{13}$CC$^{15}$N
}

\author{
B.~Tercero\inst{\ref{inst1},\ref{inst2}}
\and
N.~Marcelino\inst{\ref{inst1},\ref{inst2}}
\and
E.~Roueff\inst{\ref{inst3}}
\and
M.~Ag\'undez\inst{\ref{inst4}}
\and
C.~Cabezas\inst{\ref{inst4}}
\and
%E.~Roueff\inst{\ref{inst4}}
%\and
R.~Fuentetaja\inst{\ref{inst4}}
\and
P.~de~Vicente\inst{\ref{inst2}}
\and
J.~Cernicharo\inst{\ref{inst4}}
}

\institute{
Observatorio Astron\'omico Nacional (IGN), C/ Alfonso XII 3, 28014 Madrid, Spain\label{inst1}
\and
Observatorio de Yebes (IGN). Cerro de la Palera s/n, 19141 Yebes, Guadalajara, Spain\label{inst2} \\
\email{b.tercero@oan.es}
\and
LERMA, Observatoire de Paris, PSL Research University, CNRS, Sorbonne Universit\'e, 92190 Meudon, France\label{inst3}
\and
Grupo de Astrof\'isica Molecular, Instituto de F\'isica Fundamental, CSIC, C/ Serrano 123, 28006 Madrid, Spain\label{inst4} \\
\email{jose.cernicharo@csic.es}
}
%\and
%LOMC - UMR 6294, CNRS-Universite du Havre, 25 rue Philippe Lebon, BP1123, F-76 063 Le Havre cedex, France\label{inst4}
%\and
%Univ Rennes, CNRS, IPR (Institut de Physique de Rennes) -- UMR 6251, F-35000 Rennes, France\label{inst5}

\date{Received: 12 December 2023 / Accepted: 22 January 2024}

% \abstract{}{}{}{}{} 
% 5 {} token are mandatory 

\abstract
{We report the first detection in space of a complete sample of nine doubly substituted isotopologues of HCCCN
towards the cyanopolyyne peak of TMC-1 using observations of the QUIJOTE$^1$ line survey taken with 
the Yebes 40\,m telescope.
We detected D$^{13}$CCCN, DC$^{13}$CCN, DCC$^{13}$CN, DCCC$^{15}$N, H$^{13}$C$^{13}$CCN, H$^{13}$CC$^{13}$CN, 
HC$^{13}$C$^{13}$CN, HCC$^{13}$C$^{15}$N, and HC$^{13}$CC$^{15}$N through their $J=4-3$ and $J=5-4$ lines in the 7\,mm window.
In addition, we present an extensive analysis of the emission of HCCCN and its 
singly substituted isotopologues through a large velocity gradient model of the lines
detected at 7\,mm and 3\,mm using the Yebes 40\,m and the IRAM 30\,m telescopes, respectively.
The derived column densities for all the isotopologues are consistent in the two spectral bands 
for an H$_2$ volume density of $1\times10^4$\,cm$^{-3}$ and a kinetic temperature of 10\,K.
%These results also allow us to provide accurate isotopic ratios.
Whereas we observed a $^{13}$C fractionation for
HCC$^{13}$CN and other double isotopologues with a $^{13}$C atom adjacent to the nitrogen atom,
we derived similar C/$^{13}$C abundance ratios for the three $^{13}$C substituted species of DCCCN. This suggests 
additional chemical discrimination for deuterated isotopologues
of HCCCN.
Finally, we present the spatial distribution of the $J=4-3$ and $J=5-4$ lines from the singly substituted
species observed with the Yebes 40\,m telescope. The emission peak of the spatial distribution of DCCCN appears to be displaced by $\sim$40$''$
with respect to that of HCCCN and the $^{13}$C and $^{15}$N isotopologues. In addition to a different formation route for
the deuterated species,
we could also expect that this differentiation owing to
the deuterium fractionation is more efficient at low temperatures, and therefore, that deuterated species trace
a colder region of the cloud.}

\keywords{Astrochemistry --
                ISM: abundances --
                ISM: clouds, TMC-1 --
                ISM: molecules --
                line: identification
               }

\titlerunning{Doubly substituted isotopologues of HCCCN in TMC-1}

\maketitle
%
%-------------------------------------------------------------------

\section{Introduction}
The ultra-sensitive line survey of TMC-1 with the QUIJOTE\footnote{Q-band
Ultrasensitive Inspection Journey to the Obscure TMC-1 Environment.} project \citep{Cernicharo2021} has revealed a complex chemistry in this source, with
the detection of more than 50 new molecular species \citep[see e.g.][and references therein]
{Cernicharo2021,Cernicharo2021a,Cabezas2022,Agundez2021a,Cernicharo2023a}.
This new approach of ultra-sensitivity, \mbox{$\sigma<0.1$\,mK}, permits us to detect most of the molecules that are present in
the cloud and allows new discoveries. However, a concern connected to this approach is related to the identification of the lines
from the isotopologues of molecular species with line intensities \mbox{$\ge30$\,mK}.
This means well known molecules such as HCCCN, HCCNC, HNCCC, CH$_3$CCH, CH$_3$CN, 
C$_4$H, C$_3$H, C$_3$N, C$_6$H, and
many sulphur-bearing species (\citealt{Cernicharo2021b}, Fuentetaja et al. in prep.). Some of these species are so abundant that even doubly substituted isotopologues
are expected to be present in QUIJOTE. Many of the 2500 still unidentified features in QUIJOTE
arise from these isotopologues. In order to progress in the discovery of new species, we have undertaken
a systematic analysis of singly and doubly substituted species of these abundant molecules.

In this Letter, we report the first identification in space of a complete and coherent sample of doubly substituted species of HCCCN,
including D$^{13}$CCCN, DC$^{13}$CCN, DCC$^{13}$CN, DCCC$^{15}$N, H$^{13}$C$^{13}$CCN, H$^{13}$CC$^{13}$CN, 
HC$^{13}$C$^{13}$CN, HCC$^{13}$C$^{15}$N, and HC$^{13}$CC$^{15}$N in the dark starless cloud TMC-1 (Sect.\,\ref{results}). 
%Among these species, only DC$^{13}$CCN was already identified in the interstellar medium (ISM), in the cold dark cloud L483 \citep{Agundez2019} through the 
%detection of a single line at 84.149\,GHz ($J=10-9$). 
Of these species, H$^{13}$C$^{13}$CCN, H$^{13}$CC$^{13}$CN, and HC$^{13}$C$^{13}$CN have previously been identified in the interstellar medium 
(ISM) in the planetary nebula K4-47 \citep{Schmidt2019}, and they were tentatively detected in the high-mass star-forming region Sgr\,B2 \citep{Belloche2016}.
A thorough large velocity gradient (LVG) analysis of the singly and doubly substituted
isotopologues of HCCCN in TMC-1 is also provided by combining the emission detected at 7\,mm and 3\,mm using the data achieved with the Yebes 40\,m
and the IRAM 30\,m telescopes, respectively (Sect.\,\ref{lvg}). Then, reliable isotopic ratios in the region are derived (Sect.\,\ref{iso}).
In Sect.\,\ref{discussion} we show the spatial distribution of the $J=4-3$ and $J=5-4$ lines from the singly substituted
species observed with the Yebes 40\,m telescope and discuss the results. 
Finally, our main conclusions are summarised in Sect.\,\ref{conclusions}.

%--------------------------------------------------------------------
\section{Observations}
\label{observations}

The data presented here are part of the QUIJOTE line survey performed at the Yebes 40\,m radio 
telescope\footnote{\texttt{http://rt40m.oan.es/rt40m$\_$en.php}} \citep{Cernicharo2021,Cernicharo2023a},
 the IRAM 30\,m line survey of TMC-1 \citep{Cernicharo2012,Marcelino2023}, and the 
 imaging-line survey SANCHO\footnote{Surveying the Area of the Neighbour TMC-1 Cloud through Heterodyne Observations},
 carried out with the Yebes 40\,m telescope \citep{Cernicharo2023}.
These observations were performed towards the cyanopolyyne peak (CP) position in TMC-1 at
$\alpha_{J2000}=4^{\rm h} 41^{\rm  m} 41.9^{\rm s}$ and $\delta_{J2000}=+25^\circ 41' 27.0''$.

We covered the full Q band (7\,mm) at the 40\,m telescope, between 31.1\,GHz and 50.4\,GHz
using the NANOCOSMOS high-electron-mobility transistor (HEMT)
receiver and the fast Fourier-transform spectrometers (FFTS)
with 8$\times$2.5\,GHz bands per linear polarization,
which allow a simultaneous scan with a bandwidth of 18\,GHz at a spectral resolution of 38\,kHz \citep{TerceroF2021}.
Details of the observations and data reduction were provided by \citet{Cernicharo2021} and \citet{Cernicharo2023a}.
%The observations were performed in several sessions, between November 2019 and July 2023,
%using the frequency switching technique with a different frequency throws and setups at different central frequencies
%in order to check for spurious signals and
%other technical artefacts.
The IRAM 30\,m data consist of a 3\,mm line survey that covers the full available band at the IRAM 30\,m telescope
between 71.6 GHz and 117.6 GHz, using the EMIR 090 receiver connected to the FFTS in its narrow mode, which
provides a spectral resolution of 49\,kHz and a total bandwidth of 7.2\,GHz.
%Observations were performed in several runs. Between January and May 2012, we
%completed the scan $82.5-115.6$\,GHz, and in August 2018,
%after the upgrade of the E090 receiver, we extended the survey down to 71.6 GHz.
%Observations at IRAM 30\,m were also performed using the frequency switching technique.
Additional details were provided by \citet{Agundez2023} and \citet{Cernicharo2023a}.
Observations for the SANCHO maps were described in \citet{Cernicharo2023}.
These data were recorded
%in February 2022, December 2022, and January 2023
at the Yebes 40\,m telescope using the
on-the-fly mode in frequency switching with a throw of 10\,MHz. 

For all these data, the intensity scale in the spectra is T$_{\rm A}^*$, the antenna temperature corrected for atmospheric absorption and spillover losses, which was 
calibrated using two absorbers at different temperatures and the atmospheric transmission model 
ATM \citep{Cernicharo1985,Pardo2001}. The estimated global uncertainty for T$_{\rm A}^*$ is 10\%. 
However, the relative calibration within the Q band in the line survey and in the maps
is much better, and it is dominated by the statistical noise produced by the system.
Pointing and focus were checked every one to two hours, and the pointing errors were within 2-3$''$.
Details of the efficiencies and beam half-power intensities for the observations presented here
are given in Table\,\ref{telescopes}.
%For the Yebes 40m telescope, the main beam efficiency ranges from 0.61 at 32\,GHz to 0.47 at 49\,GHz,
%and the the half power beam width (HPBW) is 54$''$ at 32\,GHz and 36$''$ at 49\,GHz.
%At the IRAM 30m, the HPBW varies between 34$''$ and 21$''$ in the observed band,
%while the main beam efficiency varies between 0.83 at 71.6\,GHz and 0.78 at 117.6\,GHz.
All the data were reduced and analysed using the 
\texttt{GILDAS}\footnote{\texttt{http://www.iram.fr/IRAMFR/GILDAS/}} software.

%
%--------------------------------------------------------------------
\section{Results}
\label{results}

\begin{figure}
\centering
\includegraphics[width=1\columnwidth]{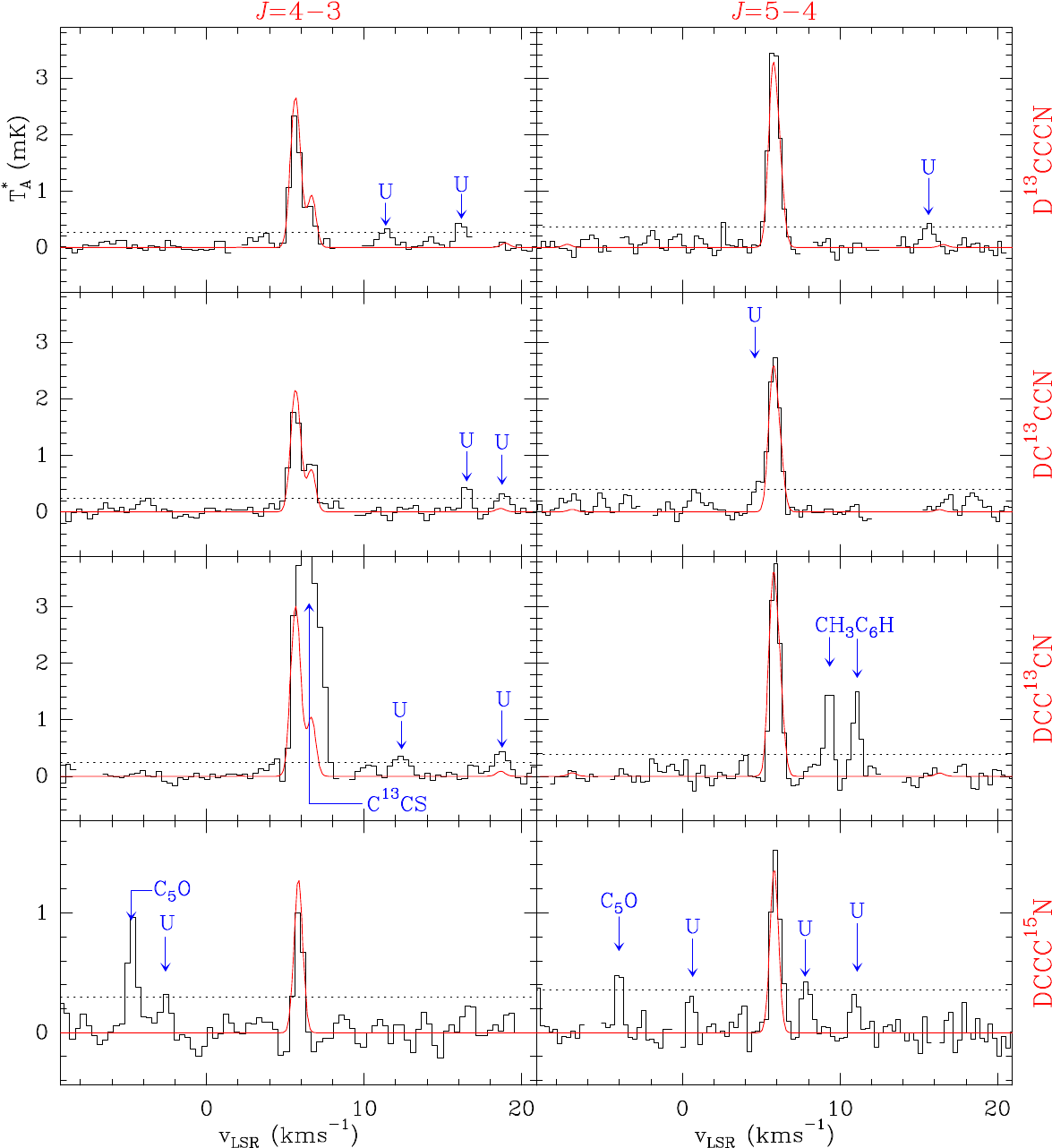}
\caption{Observed lines of 
D$^{13}$CCCN, DC$^{13}$CCN, DCC$^{13}$CN, and DCCC$^{15}$N towards TMC-1 (CP) with the Yebes 40\,m telescope (histogram black spectra). The red curves present the synthetic spectra obtained using LVG approximation (see Sect.\,\ref{lvg} and Table\,\ref{table_cd}). The dashed line represents the 3$\sigma$ level.}
\label{fig_d13c}
\end{figure}

\begin{figure}
\centering
\includegraphics[width=1\columnwidth]{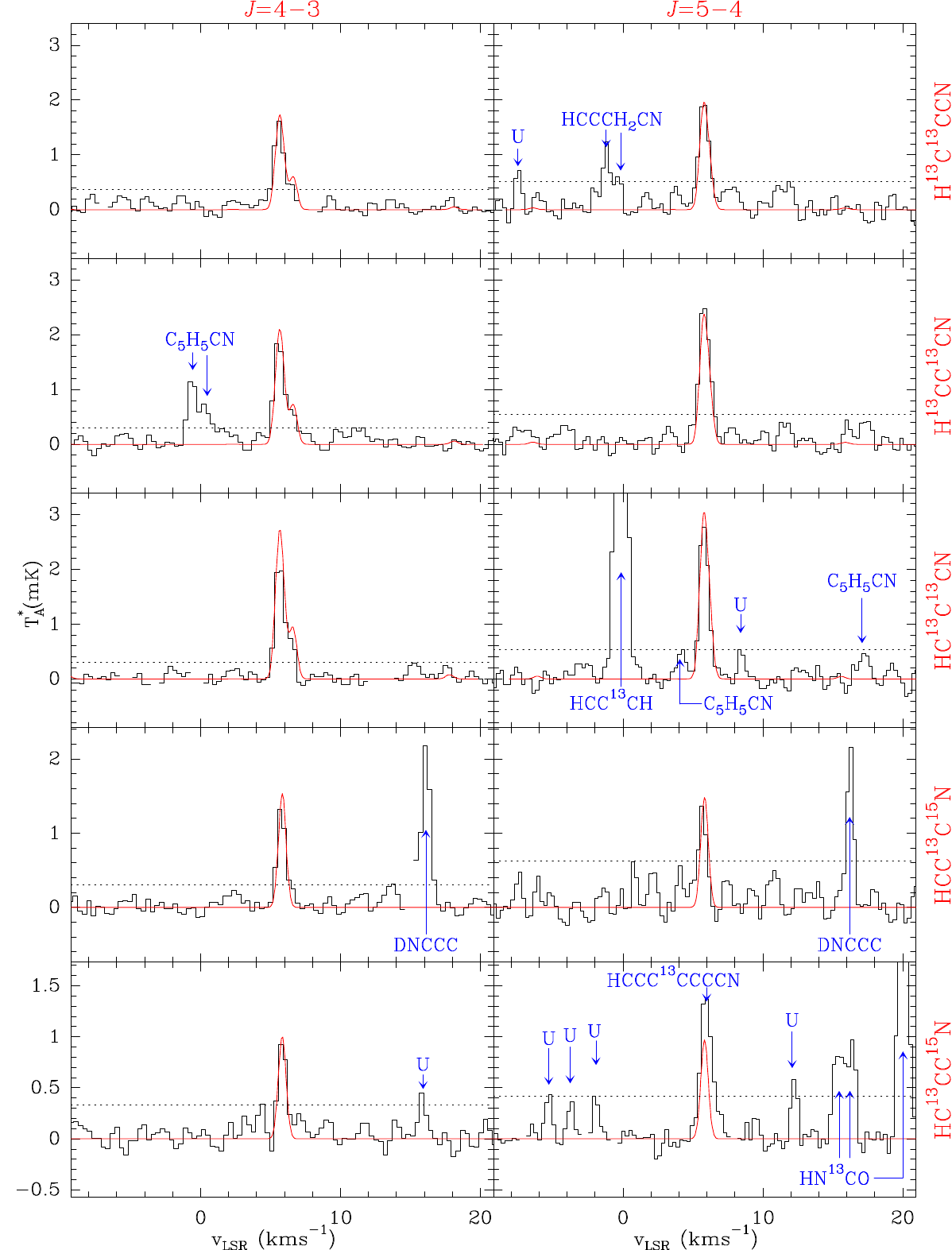}
\caption{Observed lines of
H$^{13}$C$^{13}$CCN, H$^{13}$CC$^{13}$CN, HC$^{13}$C$^{13}$CN, HCC$^{13}$C$^{15}$N, and HC$^{13}$CC$^{15}$N
towards TMC-1 (CP) with the Yebes 40\,m telescope (histogram black spectra). The red curves show the synthetic spectra obtained using LVG approximation (see Sect.\,\ref{lvg} and Table\,\ref{table_cd}). The dashed line represents the 3$\sigma$ level.}
\label{fig_13c13c}
\end{figure}

The ultra-high sensitivity of QUIJOTE allowed us to
detect the $J=4-3$ and $J=5-4$ lines of D$^{13}$CCCN, DC$^{13}$CCN, DCC$^{13}$CN, DCCC$^{15}$N, H$^{13}$C$^{13}$CCN, H$^{13}$CC$^{13}$CN, 
HC$^{13}$C$^{13}$CN, HCC$^{13}$C$^{15}$N, and HC$^{13}$CC$^{15}$N in the 7\,mm window. In addition, we identified the $J=6-5$ line of D$^{13}$CCCN and DCCC$^{15}$N
at 49.286\,GHz and 49.204\,GHz, respectively.
This complete sample of doubly substituted isotopologues of HCCCN is identified here for
the first time in space.

The observed lines are shown in Figs.\,\ref{fig_d13c} and \ref{fig_13c13c}
and are listed in Table\,\ref{tab_fits_doubleiso}.
Line identification was performed using the catalogue provided by \texttt{MADEX} (\citealt{MADEX}; see Table\,\ref{tab_fits_doubleiso}).
An extensive set of laboratory data is available for all these species, except for 
H$^{13}$CCC$^{15}$N and HC$^{13}$CC$^{15}$N.
The laboratory frequencies are taken from \citet{deZafra1971,Creswell1977,Mallinson1978,Plummer1988,Chen1991,Yamada1995,Mbosei2000,
Thorwirth2000,Thorwirth2001,Spahn2008}. We fitted these frequencies with a standard Hamiltonian for linear
molecules with hyperfine structure to derive their rotational constants. They were implemented in
the \texttt{MADEX} code, providing the frequencies shown in Table\,\ref{tab_fits_doubleiso}.
To determine the rest frequencies of 
H$^{13}$CCC$^{15}$N and
HC$^{13}$CC$^{15}$N, we predicted their rotational constants from the substitution structure of HCCCN derived from the rotational constants of all its isotopologues
measured in the laboratory.
One pair of lines was easily found in perfect harmonic relation 5/4 within a few MHz of the predicted frequencies for
HC$^{13}$CC$^{15}$N.
However, no such pair of lines could be unambiguously identified for H$^{13}$CCC$^{15}$N because it was
blended with other strong features.
Adopting for HC$^{13}$CC$^{15}$N a distortion constant interpolated between those of the other isotopologues 
($D=0.509$\,kHz), we derive $B$(HC$^{13}$CC$^{15}$N)\,=\,4396.6604(5)\,MHz with a standard deviation of the fit of 1.7\,kHz.
%Spectroscopic line parameters were obtained using \texttt{MADEX} by fitting all the rotational lines reported by \citet{Spahn2008} and
%\citet{Thorwirth2001}. For HC$^{13}$CC$^{15}$N we fit the rotational lines observed with QUIJOTE (see Sect.\,\ref{results}).
%Dipole moment from \citet{DeLeon1985}.
A single-Gaussian function was fitted to the line profiles
to obtain the observed line parameters (see Table\,\ref{tab_fits_doubleiso}).
The derived radial velocities ($v_{\rm LSR}$) and line widths ($\Delta$$v$; full width at half maximum, FWHM)
agree with the mean values obtained by \citet{Cernicharo2020a} from 
Gaussian fits to the 50 lines of HC$_5$N and its $^{13}$C and $^{15}$N isotopologues 
detected in this survey. 
Lines from all the singly substituted isotopologues of HCCCN and from the parent species
are identified in the two spectral windows, at 7\,mm and 3\,mm.
Figures \ref{fig_normal_simple_7mm}, \ref{fig_normal_3mm}, and \ref{fig_simple_3mm} show these detections.
The observed line parameters derived from Gaussian fits are listed in Tables\,\ref{tab_fits_normal} and \ref{tab_fits_single}.

\subsection{LVG analysis}
\label{lvg}

To derive molecular column densities and constrain the H$_2$ volume density in the source,
we performed a combined LVG analysis of HCCCN and its isotopologues at 7\,mm and 3\,mm.
For all the studied species, we used the collisional cross sections between HCCCN and para-H$_2$ provided by \citet{Faure2016}.
We assumed that the main collider is $p$-H$_2$, as expected in cold dark clouds (see e.g. \citealt{Flower2006}).

Firstly, we used the weak hyperfine satellite lines ($S_{\rm r}$, $S_{\rm w}$, and $S_{\rm b}$\footnote{$S_{\rm r}$: $F_{\rm u}-F_{\rm l}=J-J$; $S_{\rm w}$: $F_{\rm u}-F_{\rm l}=(J-1)-J$; $S_{\rm b}$: $F_{\rm u}-F_{\rm l}=(J-1)-(J-1)$.})
of the $J=4-3$ and $J=5-4$ transitions of HCCCN  
and the $J=4-3$ and $J=5-4$ lines of the singly substituted species (see Fig.\,\ref{fig_normal_simple_7mm}) to estimate
initial column densities. We assumed a source diameter of 80$''$ \citep{Fosse2001},
a radial velocity of 5.83\,km\,s$^{-1}$, an FWHM of 0.6\,km\,s$^{-1}$, and 
an excitation temperature of 10\,K in agreement
with the excitation temperatures derived for the HCCCN isomers (see \citealt{Cernicharo2020a})
and the kinetic temperature of the source, which was derived using
NH$_3$ and other symmetric top molecules (see e.g. \citealt{Feher2016,Agundez2023}).
It is worth noting that 
these lines are not too sensitive
to the excitation temperature as soon as it is higher than $6-7$\,K.

Using these priors, we fitted the density that reproduces the 3\,mm lines of the singly substituted
isotopologues (Fig.\,\ref{fig_simple_3mm}) obtaining a result of $n$(H$_2$)\,$=1\times10^4$\,cm$^{-3}$.
This value was also
derived by \citet{Agundez2023} using the singly susbtituted $^{13}$C isotopologues of HC$_3$N.
Using this volume density, we fitted all the lines of the
singly substituted isotopologues at 7\,mm and 3\,mm with a single column density for each species (Figs.\,\ref{fig_normal_simple_7mm} and \ref{fig_simple_3mm}).
To obtain a reasonable fit for the HCCCN lines at 3\,mm,
we increased the hydrogen volume density to $n$(H$_2$)\,$=2\times10^4$\,cm$^{-3}$ (Fig.\,\ref{fig_normal_3mm}).
However, this change in the density arises because the HCCCN lines are optically thick.
%and that some intensity anomalies appear between
%the $S_{\rm r}$ and $S_{\rm b}$ hyperfine components in the lines from $J=8-7$ to $J=12-11$.
Consequently, we derived the column density of the 
doubly substituted isotopologues, detected only at 7\,mm, assuming the LVG formalism as well, $n$(H$_2$)\,$=1\times10^4$\,cm$^{-3}$, $T_{\rm K}=10$\,K,
and the line parameters given above.

The final synthetic spectra obtained from these models reproduce the observed line profiles rather well.
Comparisons between model and observation are shown in Figs.\,\ref{fig_d13c}, \ref{fig_13c13c},
\ref{fig_normal_simple_7mm}, \ref{fig_normal_3mm}, and \ref{fig_simple_3mm}.
The derived column density values for each species are listed in Table\,\ref{table_cd}. 
Owing to the good fits that modelled a large number of observed lines at different frequency domains,
%\textcolor{blue}{(comentario: Yo asignaría el error estadistico =+-sigma/intensidad)}
and the high signal-to-noise ratio of the observed lines, we assigned as uncertainty to the column density the statistical error
derived from error propagation of the root mean square error of the data.

To evaluate the rotational temperature ($T_{\rm rot}$), and thus the level of departure from local thermodynamic equilibrium (LTE),
and for an independent estimate of the column density ($N$), we also constructed rotation diagrams (see e.g. \citealt{Goldsmith1999})
for all species detected at both 3\,mm and 7\,mm (HCCCN, DCCCN, H$^{13}$CCCN, HC$^{13}$CCN, HCC$^{13}$CN, and HCCC$^{15}$N) for which a large number of lines are detected (see Appendix\,\ref{lte} and 
Table\,\ref{table_cd}).

\subsection{Isotopic ratios}
\label{iso}

From the LVG column densities reported above, we derived the isotopic
abundance ratios provided in Table\,\ref{table_cd}. 

%%%%% evelyne

A significant number of deuterated molecules has already been detected in the TMC-1 environment \citep{Cabezas2021a,Cabezas2021b,Cabezas2022a,Navarro-Almaida2021}.
%including doubly deuterated species as
%reported in \citet{Agundez2021b}.
%Owing to the low temperature of the source we expect an enhancement of the abundance of heavy isotopologues
%due to the lowering of the zero-point vibrational energy, specially for
%for the deuterated species. This
%energy  difference explains the exothermicity in cold gas of reactions such as
%\ce{ H3+ + HD <=> H2D+ + H2 } or
%\ce{ HCN + HD <=> DCN + H2 } \citep{Solomon1973}.
%Further deuteration may proceed via formation of D$_2$H$^+$ and D$_3$$^+$ in reactions with HD and D$_2$.
%The deuterated ions also react with a variety of neutral molecules leading to high abundances
%of many deuterated species (see, e.g., \citealt{Albertsson2013}).
The deuteration fraction is indeed enhanced by several orders of magnitude in cold clouds
compared to the elemental D/H ratio of $\sim1.5\times10^{-5}$
\citep{Linsky2003} as a result of the
%significant energy difference compared to the ambient temperature ($\sim$ 192\,K versus 10\,K) involved in the
significant exothermicity of the reaction \ce{ H3+ + HD <=> H2D+ + H2 } (192\,K),
%deuterium isotopic exchange reaction,
as first pointed out by \citet{Watson1976}.

Isotopic exchange reactions may take place for $^{13}$C, $^{18}$O, and $^{15}$N containing species as well, as suggested first by \citet{Langer1984} and \citet{Terzieva2000}
and updated later in complex chemical models \citep{Rodgers2008,Roueff2015,Wirstrom2018,Loison2019,Loison2020,Sipila2023}.
However, the zero-point energy variation is much smaller
for the heavier atoms $^{13}$C and $^{15}$N, which induces much lower isotopic enhancements.

%\textcolor{blue}{(Comentario: la energía de enlance, binding, no depende de la masa
%de los núcleos, sólo de la carga. El ZPE depende de las vibraciones, es decir, de la masa de los
%núcleos.)}
%\begin{itemize}
%\item`

%\textit{Deuterium substitution:}

For deuterated molecules, we detected both DC$_3$N and the different D containing isotopologues with one $^{13}$C atom and one $^{15}$N isotopic substitution.
The D/H ratios of these different isotopologues are
very similar, between 0.013 and 0.018, in agreement with the ratio
found by \citet{Turner2001} and \citet{Cernicharo2020a} for DCCCN/HCCCN in the source.
%From the deuterated species, we find a D/H ratio between 0.013 and 0.018s
%for the three $^{13}$C substitutions in HCCCN and DCCCN, in agreement with that
%found by \citet{Turner2001,Cernicharo2020a} for DCCCN/HCCCN in the source.
%We note that the deuterium fractionation ratios in TMC-1 tend to be systematically lower than those
%of other low-mass prestellar and protostellar sources except for the ratio
%derived from CH$_3$OD (see Table\,3 of \citealt{Agundez2019}).
HCCCN and DCCCN are mainly formed via the \ce{C$_2$H$_2$ + CN}, \ce{C$_4$H + N} and \ce{C$_2$HD  + CN}, \ce{C$_4$D + N} reactions, respectively,
and from the dissociative recombination of HC$_3$NH$^+$ and DC$_3$NH$^+$. The molecular ion DC$_3$NH$^+$ is enriched in deuterium via the
\ce{HCCCN + H$_2$D$^+$} reaction.
The DCCCN/HCCCN ratio is then slightly higher than that of C$_4$D/C$_4$H, which was found to be 0.008 \citep{Cabezas2021b}.
Similar reactions are most probably involved in the $^{13}$C substituted species.

%\textit{$^{13}$C substitutions:}

We detect singly and doubly $^{13}$C substituted HCCCN as well as $^{13}$C plus $^{15}$N substitutions with different locations of $^{13}$C.
The C/$^{13}$C ratios derived from the different isotopologue substitutions of HCCCN
show a clear fractionation for the species with the $^{13}$C isotope adjacent to the nitrogen atom that corresponds to the most stable system as
shown in \cite{Takano1998}. Under thermodynamic equilibrium, HCC$^{13}$CN is predicted to be more abundant by a factor of 2 than HC$^{13}$CCN  at 10\,K \citep{Wolfsberg1979}.
The same factor of 2 is predicted for the HC$^{13}$CCN/H$^{13}$CCCN ratio at the same temperature, following the respective zero-point energy values.
However, this feature is not seen in the observations. Kinetic effects rather than thermodynamic considerations are indeed at work
under interstellar conditions.

%For the $^{13}$C isotopologues that do not show abundance enhancement,
Most of the derived C/$^{13}$C abundance ratios are $\sim$95 and $\sim$105, consistent with but slightly higher than the Solar System
value of 89 \citep{Wilson1994}. The standard value of $\sim70$ in the local ISM (see e.g. \citealt{Wilson1994,Ritchey2011})
is similar to the values between 55 and 80 that we found for the enhanced species, that is, the
carbon near nitrogen.
The C/$^{13}$C ratio in deuterated cyanoacetylene is close to 100, with a slightly lower value for the most stable DCC$^{13}$CN isotopologue and
rising up to 132 for DC$^{13}$CCN. This could indicate a light dilution of the DC$^{13}$CCN abundances with respect the other $^{13}$C isotopologues of DCCCN.
The relative abundance ratios are 1.0:1.1:1.6 for H$^{13}$CCCN:HC$^{13}$CCN:HCC$^{13}$CN,
similar to the ratios reported previously in TMC-1 \citep{Takano1998}, L1527 \citep{Araki2016}, and L1521B \citep{Taniguchi2017}.
On the other hand, the $^{13}$C isotopologues of the deuterated species show relative abundances of 1.3:1.0:1.4 for
D$^{13}$CCCN:DC$^{13}$CCN:DCC$^{13}$CN,
which indicate different enhancement routes for 
%{\st{$^{13}$C$_3$}} 
$^{13}$C
in HCCCN and DCCCN. 
%{\sout{We note that the isotopologues $^{13}$C of HCCNC and HNCCC are also in an abundance ratio close to those derived here from DCCCN (Cernicharo et al. 2023b).}} 
We note that the three $^{13}$C isotopologues of HCCNC and HNCCC are in an abundance ratio close to $\sim$95, and therefore,
these HCCCN isomers do not present fractionation depending on the position of the $^{13}$C atom 
\citep{Cernicharo2024}.

%The relative abundance ratios are 1.0:1.1:1.6 for H$^{13}$CCCN:HC$^{13}$CCN:HCC$^{13}$CN,
%similar to the ratios reported previously in TMC-1 \citep{Takano1998}, L1527 \citep{Araki2016}, and L1521B \citep{Taniguchi2017}.
It is worth noting that for cold sources, the relative abundance ratios of the $^{13}$C isotopologues of HC$_3$N
is not dependent on the evolutionary state of the source. The abundances in L1527, however, which is in an evolved evolutionary state (protostellar core),
are similar to those of the starless cores TMC-1 and L1521B. Other cold cores present different relative abundances:
1.0:1.1:1.2 for L483 (protostellar core; \citealt{Agundez2019}) or 1.5:1.0:2.1 for L134N
(starless core; \citealt{Taniguchi2017}). All these sources present kinetic temperatures between 9 and 14\,K \citep{Agundez2023}.

\citet{Takano1998} pointed out that the HCC$^{13}$CN fractionation is consistent with
C$_2$H$_2$ + CN reaction as the main formation route for HCCCN when we assume that the triple bond of C$_2$H$_2$ is
conserved without breaking in the reaction (and then the abundances of H$^{13}$CCCN and HC$^{13}$CCN
remain equal to each other) and the CN molecule introduces the $^{13}$C enriched carbon into HCCCN. The observed fractionation also excludes isotope exchange with HCCCN as the main process
for the formation of the $^{13}$C isotopologues.

For the double isotopologues, we also found an enhancement of H$^{13}$CC$^{13}$CN and HC$^{13}$C$^{13}$CN relative to H$^{13}$C$^{13}$CCN and of HCC$^{13}$C$^{15}$N relative to HC$^{13}$CC$^{15}$N,
which points to a fractionation for all the non-deuterated double isotopologues in which the $^{13}$C isotope is adjacent to the nitrogen atom. For these cases, isotope exchange with HCC$^{13}$CN might
proceed as an important route for the formation of these double isotopologues.

For the $^{15}$N substitutions, the isotopic ratios of the hydrogen and deuterated cyano-acetylene are very similar, with respective values of 317 and 330.
The values derived for the $^{13}$C containing $^{15}$N isotopes are slightly lower, equal to $\sim$ 250 and independent of the $^{13}$C position.
The measured N/$^{15}$N ratios in the local ISM differ by up to a factor of two (see e.g. \citealt{Agundez2019}), with values in the
range of $237-450$. Our derived abundance ratios, between 250 and 330, are consistent with the local ISM values.
These results also agree with previous values found in TMC-1 \citep{TaniguchiSaito2017} and in other cold cores \citep{Agundez2019,Magalhaes2018,Hily-Blant2018}.

\section{Discussion}
\label{discussion}

The results of Sect.\,\ref{iso} show a $^{13}$C fractionation
for all the simple and double isotopologues in which 
$^{13}$C is adjacent to the nitrogen atom, except for DCC$^{13}$CN, which suggests
different enhancement routes for $^{13}$C$_3$ 
in HCCCN and DCCCN and may indicate different formation routes for HCCCN and DCCCN.
To further explore this hypothesis, we exploited the SANCHO data
to produce maps of the spatial distribution of HCCCN and all its singly substituted isotopologues (Fig.\,\ref{fig_map_hc3n}).
All data within a circle of 20$''$ were averaged spatially with equal
weight. At low frequency, this represents a small degradation of the
spatial resolution. However, at the highest frequency in the Q band,
the effective angular resolution is $\sim$\,1.4 higher than the half-power beam width (HPBW)
of the telescope and produces a broadening of the spatial structures.
Nevertheless, this results in an effective spatial resolution that is roughly
identical in the Q band.
A detailed description of the data reduction procedure was presented in \citet{Cernicharo2023}.
The HCCCN map was obtained
using the integrated intensity of the weak satellite hyperfine lines ($S_{\rm r}$, $S_{\rm w}$, and $S_{\rm b}$).
This guarantees low line opacities, and therefore, the integrated intensities are
proportional to the column densities. The maps of DCCCN, H$^{13}$CCCN, HC$^{13}$CCN, and HCC$^{13}$CN
correspond to the sum of the integrated intensity of all hyperfine lines for each transition.

As shown in Fig.\,\ref{fig_map_hc3n}, the integrated emission of all the isotopologues follows 
the well-known filamentary structure of TMC-1 with a dense condensation around the central position.
The SANCHO maps show that the emission peak of the integrated intensity of HCCCN and other
cyanopolyynes (see also \citealt{Cernicharo2023}) is displaced with respect to the coordinates
of our QUIJOTE survey, which is widely assumed as the CP of TMC-1 (see e.g. \citealt{Kaifu2004,McGuire2020}).
However, whereas the intensity peak is located at a south-east position from the 
centre of the map (the CP peak) for HCCCN, H$^{13}$CCCN, HC$^{13}$CCN, HCC$^{13}$CN, and HCCC$^{15}$N, 
the deuterated isotopologue emission peaks at a significantly different position,
north-east of the CP peak. These two peaks are displaced by about $\sim$40$''$.
Figure\,\ref{fig_map_hc3n} therefore shows that DCCCN alone does not present a clear co-emission
with the main isotopologue. 
This finding also points to different formations routes for the 
deuterated and non-deuterated isotopologues of HCCCN.
In addition, because deuterium fractionation is more efficient at low temperatures,
we could also expect that DCCCN traces a colder or denser region of the cloud.
In this context, we note that the DCCCN map resembles the two HC$_7$N maps
presented in \citet{Cernicharo2023}, which were performed with lines of high-energy transitions
($J=28-27$ at $E_{\rm upp}=22.0$\,K and $J=38-37$ at $E_{\rm upp}=40.1$\,K).
Although the high signal-to-noise ratio of the HC$_7$N lines provides good-quality maps,
these high-energy transitions are very sensitive to the density or to
small changes in the kinetic temperature. We
therefore cannot compare this directly with the low-energy DCCCN emission.
This prevents us from drawing stronger conclusions about the temperature or density structure of the source.

%\begin{figure*}
%\centering
%\includegraphics[width=1\textwidth]{iso_main_ratios_normalized.pdf}
%\caption{Color plot of the spatial distribution of the integrated intensity (between 5.3 km\,s$^{-1}$ and 6.5 km\,s$^{-1}$) ratios of HC$_3$N and
%its single substituted isotopologues detected in the SANCHO survey. The sampling of the data is 20$''$ and the integrating area in
%each position corresponds to a circle of 20$''$ of radius. For each species the integrated intensity
%has been normalized to the maximum value within
%the area covered by the map (see Fig.\,\ref{fig_map_hc3n}). Hence, the color scale is the same for all species and is indicated by the wedge at the lower right part
%of the Figure. The HPBW of the telescope is indicated at the bottom-right corner of each map.
%The black dot indicates the centre of the map, which corresponds to the QUIJOTE survey position, i.e. the cyanopolyyne peak of TMC-1.
%}
%\label{fig_map_ratios}
%\end{figure*}

%--------------------------------------------------------------------
\section{Conclusions}
\label{conclusions}

The QUIJOTE survey of TMC-1 allowed us to identify a complete and coherent sample of doubly substituted isotopologues of HCCCN in space for the first time.
We detected D$^{13}$CCCN, DC$^{13}$CCN, DCC$^{13}$CN, DCCC$^{15}$N, H$^{13}$C$^{13}$CCN, H$^{13}$CC$^{13}$CN, 
HC$^{13}$C$^{13}$CN, HCC$^{13}$C$^{15}$N, and HC$^{13}$CC$^{15}$N through their $J=4-3$ and $J=5-4$ lines in the 7\,mm window.
We derived column densities for all the singly and doubly substituted isotopologues and physical properties of the source using LVG approximation.
For HCCCN and its singly substituted isotopologues, we found consistent values in the two observed spectral
bands, 7\,mm and 3\,mm, and with $n$(H$_2$)\,=\,$1\times10^4$\,cm$^{-3}$ and $T_{\rm}=10$\,K.
The column density ratios between different isotopologues show a slight $^{13}$C fractionation for HCC$^{13}$CN and other doubly substituted isotopologues
in which a $^{13}$C atom lies adjacent to the nitrogen atom. This is not the case for DCC$^{13}$CN, which points to
additional chemical discrimination for deuterated isotopologues
of HCCCN.
This hypothesis is supported by the SANCHO data, which allow the comparison of the spatial distribution of HCCCN and that of its singly substituted isotopologues.
The emission peak of the spatial distribution of the deuterated species appears to be displaced by $\sim$40$''$
with respect to that of the remaining singly substituted isotopologues, suggesting different
enhancement routes for the different isotopologues of HCCCN.
%formation routes and/or
%different physical properties traced by each species.

%--------------------------------------------------------------------
\begin{acknowledgements}
We thank the anonymous
referee for helpful comments and suggestions
that improved the presentation of this paper
significantly.
We acknowledge funding support from the European Research Council (ERC Synergy Grant 610256: NANOCOSMOS). 
We also thank the Spanish MICIN for funding support under grants PID2019-106110GB-I00, PID2019-107115GB-C21, and PID2019-106235GB-I00.
This research has been also funded by grant PID2022-137980NB-I00 by the Spanish Ministry of Science and 
Innovation/State Agency of Research MCIN/AEI/10.13039/501100011033 and by “ERDF A way of making Europe”.
\end{acknowledgements}


\bibliographystyle{aa}
\bibliography{references_hc3n_evelyn}

\begin{appendix}
\section{Additional figures and tables}
\label{appen_figsTables}

Additional figures and tables are shown in this appendix. 
Table\,\ref{telescopes} shows details of the efficiencies and HPBW
for the Yebes 40\,m and the IRAM 30\,m telescopes (see Sect.\,\ref{observations}).
Observational and spectroscopic line parameters for the detected lines (see Sect.\,\ref{results}) and the excitation temperature
derived for each transition using the LVG formalism (see Appendix\,\ref{lte}) are
shown in Tables\,\ref{tab_fits_doubleiso}, \ref{tab_fits_normal}, and \ref{tab_fits_single}.
Table\,\ref{table_cd} shows the derived column density values for each species
using the LVG approximation (see Sect.\,\ref{lvg}) and the rotational diagram procedure (see Appendix\,\ref{lte}),
and the derived isotopic abundance ratios (see Sect.\,\ref{iso}).

\begin{figure}
\centering
\includegraphics[width=1\columnwidth]{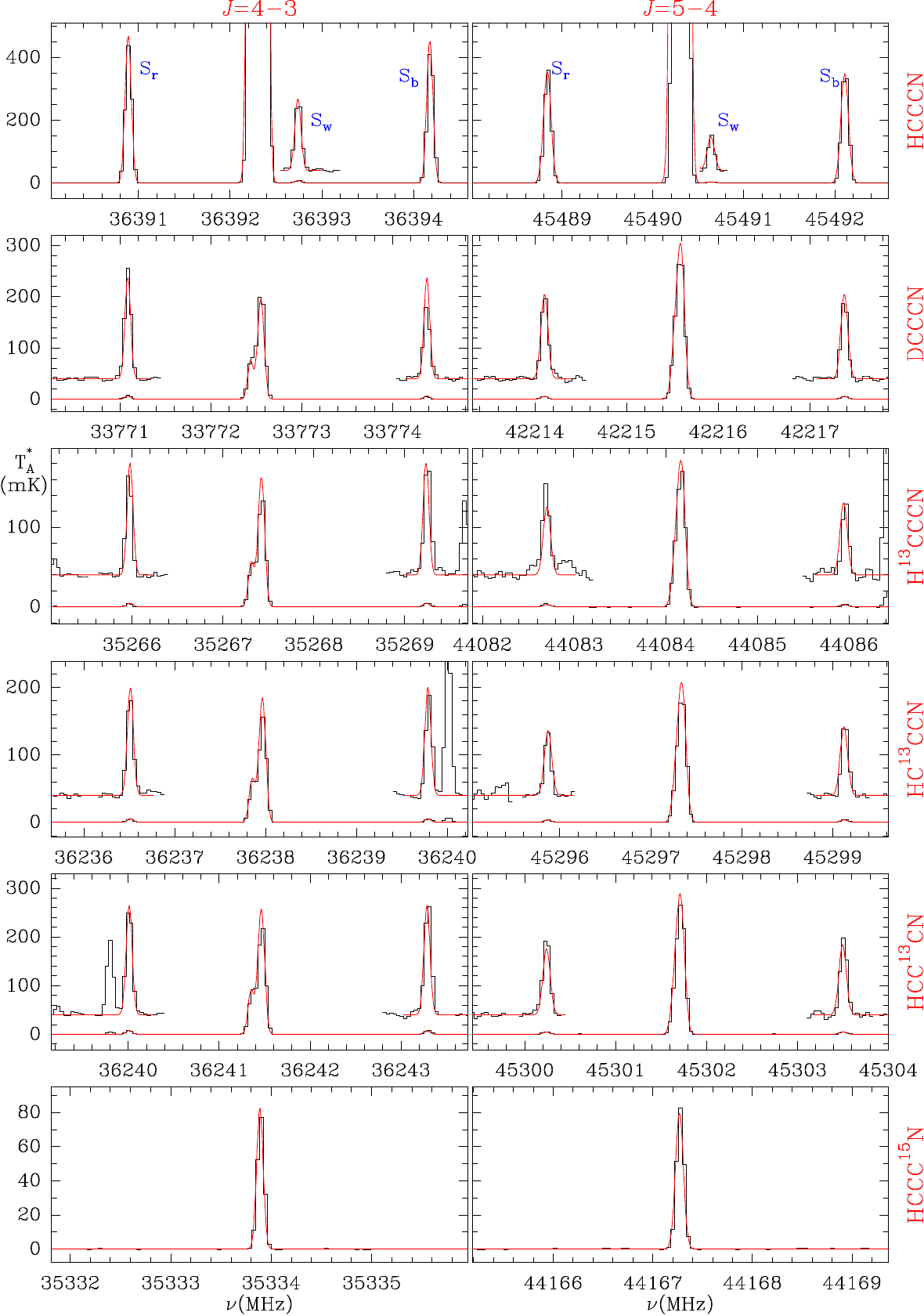}
\caption{Observed lines of 
HCCCN, DCCCN, H$^{13}$CCCN, HC$^{13}$CCN, HCC$^{13}$CN, and HCCC$^{15}$N towards TMC-1 (CP) at 7\,mm (histogram black spectra).
The red curves show the synthetic spectra obtained using the LVG approximation (see Sect.\,\ref{lvg} and Table\,\ref{table_cd}).
The weak satellite hyperfine lines ($S_{\rm w}$ for HCCCN and $S_{\rm r}$ and $S_{\rm b}$ for the isotopologues) have been offseted and multiplied by a factor of 30.
These hyperfine lines correspond to the transitions $F_{\rm u}-F_{\rm l}=J-J$ for $S_{\rm r}$, $F_{\rm u}-F_{\rm l}=(J-1)-J$ for $S_{\rm w}$, and $F_{\rm u}-F_{\rm l}=(J-1)-(J-1)$
for $S_{\rm b}$.}
\label{fig_normal_simple_7mm}
\end{figure}

\begin{figure}
\centering
\includegraphics[width=.8\columnwidth]{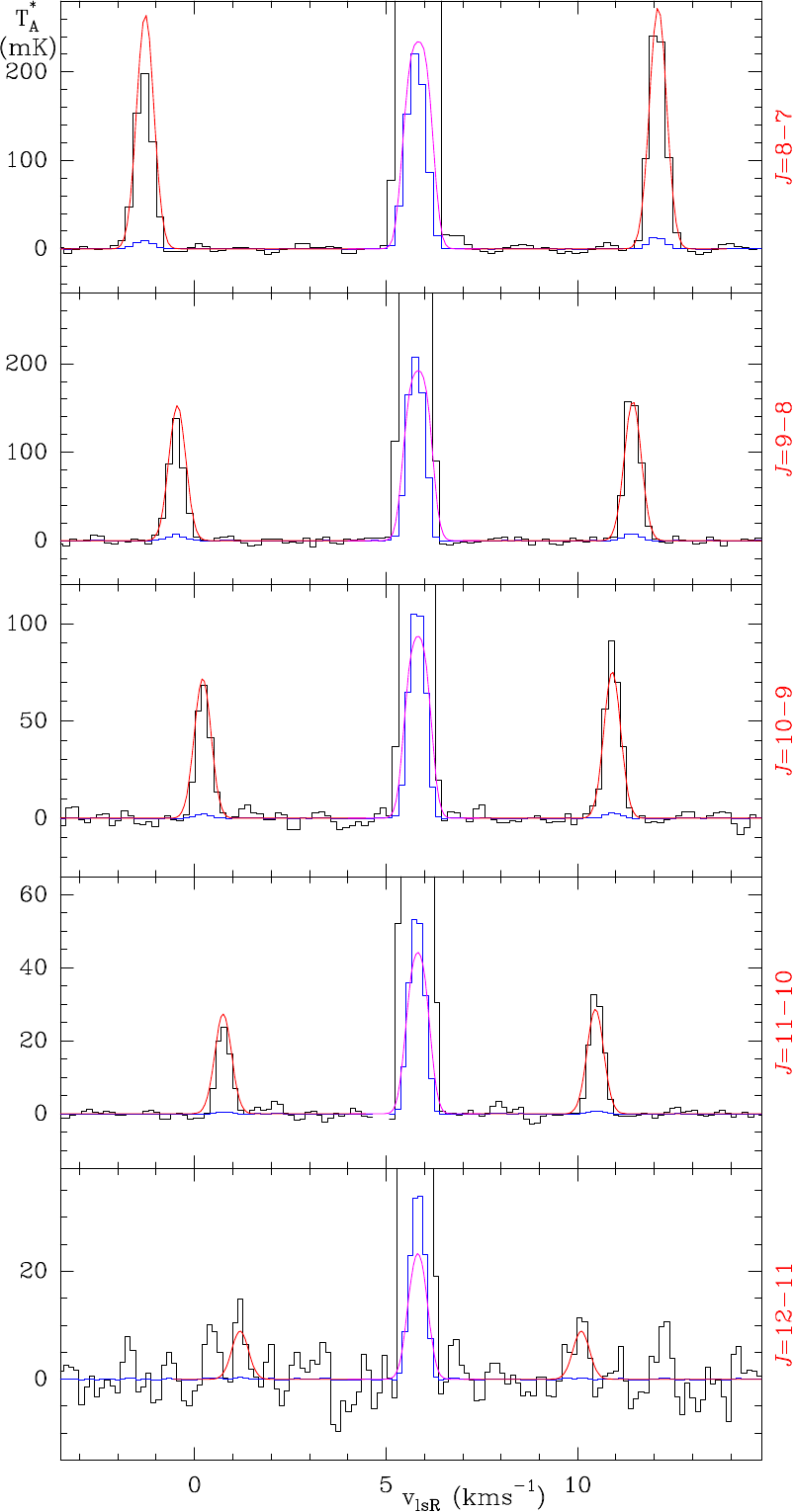}
\caption{Observed lines of
HCCCN towards TMC-1 (CP) at 3\,mm (histogram black spectra). The blue histogram spectra show the data divided by a factor of 20, 20, 30, 40, and 40, from top to bottom.
The red curves show the synthetic spectra obtained using the LVG approximation (see Sect.\,\ref{lvg} and Table\,\ref{table_cd}). The magenta curves show the synthetic spectra divided
by the same factors as above.}
\label{fig_normal_3mm}
\end{figure}

\begin{figure}
\centering
\includegraphics[width=1\columnwidth]{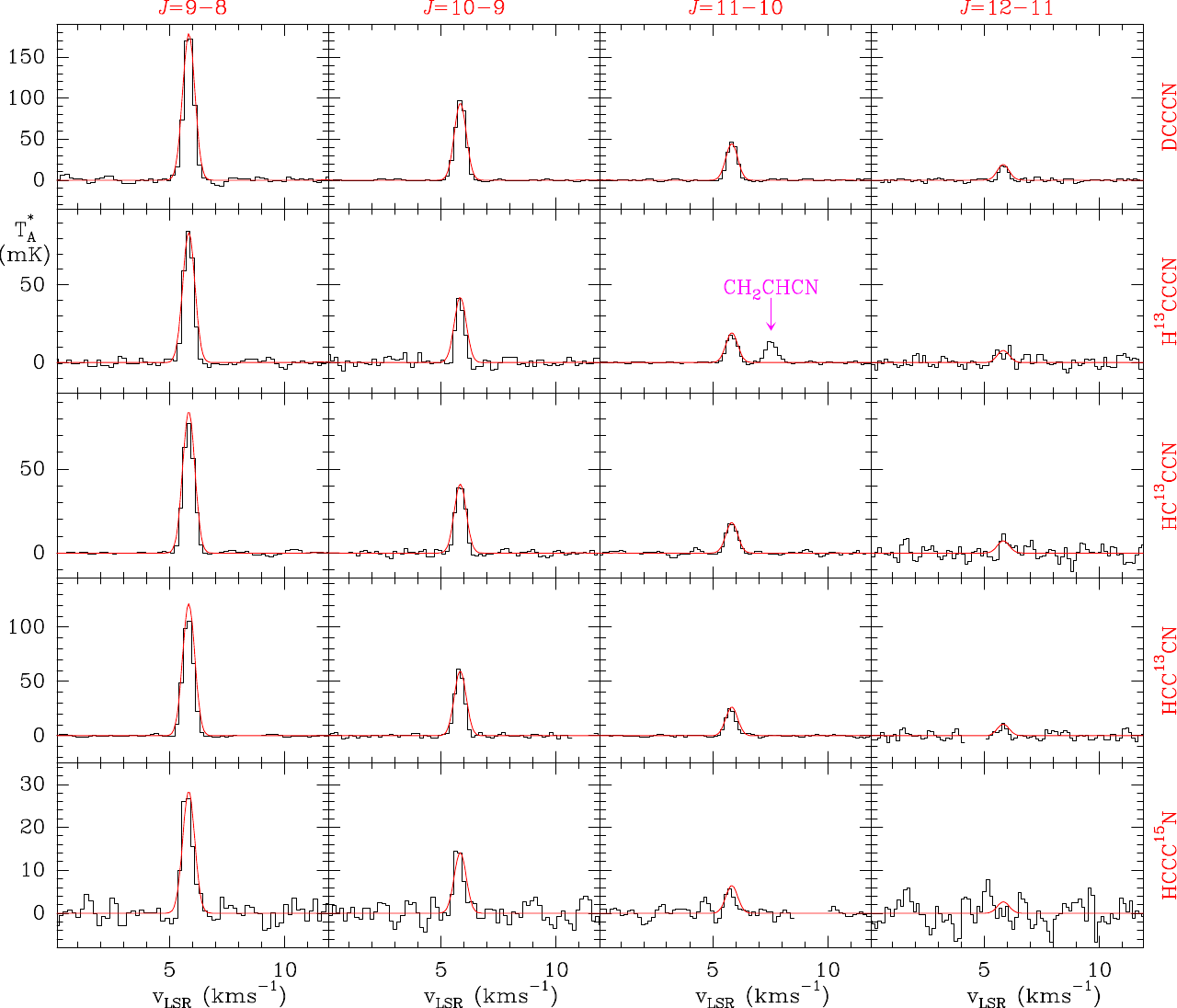}
\caption{Observed lines of
DCCCN, H$^{13}$CCCN, HC$^{13}$CCN, HCC$^{13}$CN, and HCCC$^{15}$N towards TMC-1 (CP) at 3\,mm (histogram black spectra). The red curves show the synthetic spectra obtained using the LVG approximation (see Sect.\,\ref{lvg} and Table\,\ref{table_cd}).}
\label{fig_simple_3mm}
\end{figure}

\begin{figure*}
\centering
\includegraphics[width=1\textwidth]{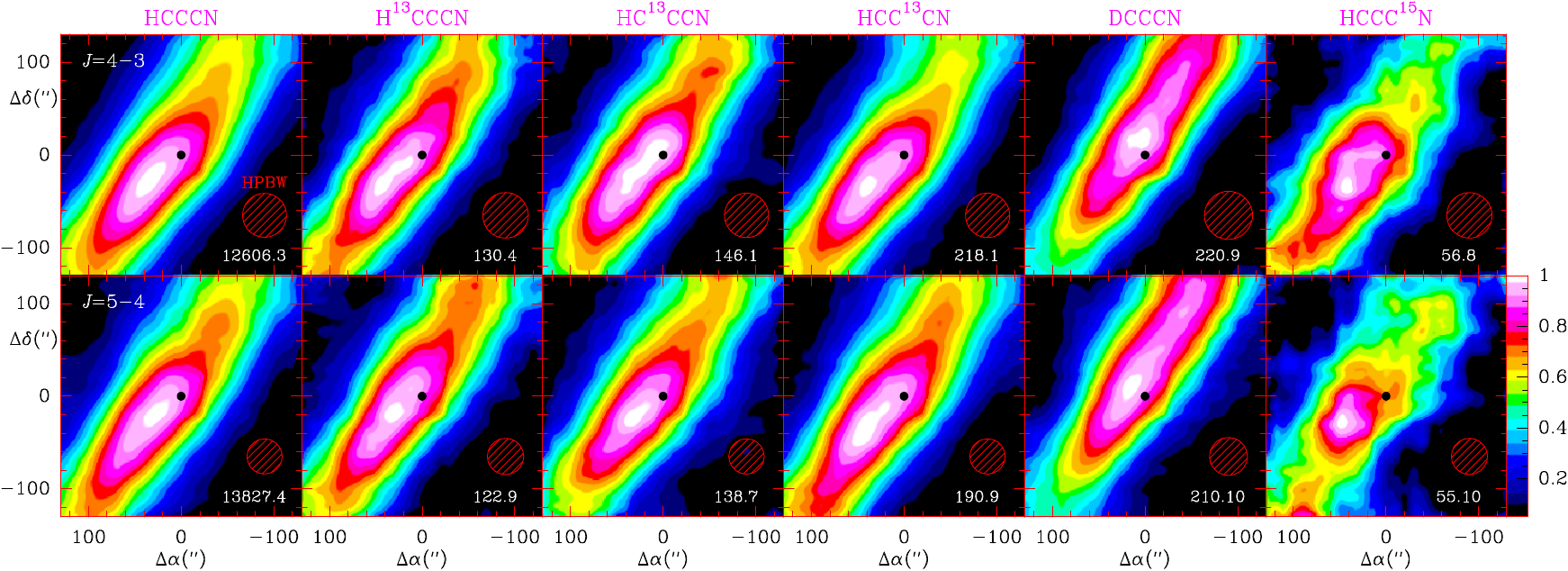}
\caption{Colour plot of the spatial distribution of the integrated intensity
between 5.3 km\,s$^{-1}$ and 6.5 km\,s$^{-1}$ of the $J=4-3$ and $J=5-4$ lines of HCCCN and
its singly substituted isotopologues detected in the SANCHO survey. The sampling of the data is 20$''$, and the integrating area in
each position corresponds to a circle with a radius of 20$''$ (see Sect.\,\ref{discussion}). For each species, the integrated intensity
has been normalized to the maximum value within
the area covered by the map. The colour scale is therefore the same for all species and is indicated by the wedge in the lower right part
of the figure. The maximum intensities (in mK\,km\,s$^{-1}$) are indicated in the bottom right corner of each panel.
The HPBW of the telescope is also indicated in the bottom right corner of each map.
The black dot indicates the centre of the map, which corresponds to the QUIJOTE survey position, i.e. the cyanopolyyne peak of TMC-1.
}
\label{fig_map_hc3n}
\end{figure*}

The detected lines from all the singly substituted isotopologues of HCCCN and 
from the parent species (see Sect.\,\ref{results}) and the synthetic spectra
derived using the LVG approximation (see Sect.\,\ref{lvg}) are shown in 
Figs.\,\ref{fig_normal_simple_7mm}, \ref{fig_normal_3mm}, and \ref{fig_simple_3mm}.
Figure\,\ref{fig_map_hc3n} shows the spatial distribution of the integrated intensity
between 5.3 km\,s$^{-1}$ and 6.5 km\,s$^{-1}$ of the $J=4-3$ and $J=5-4$ lines of HCCCN and
its singly substituted isotopologues detected in the SANCHO survey.

\begin{table}[tbh]
\begin{center}
\caption[]{Half-power beam width, main-beam efficiency, and forward efficiency
of the Yebes 40\,m and IRAM 30\,m radiotelescopes in the Q and W bands, respectively.}
\label{telescopes}
\begin{tabular}{rcccc}
\hline \noalign {\smallskip}
Frequency & Wavelength & HPBW & $\eta_{\rm MB}$ & $\eta_{\rm F}$ \\
(GHz) & (mm) & ($''$)  & & \\
\hline \noalign {\smallskip}
\hline \noalign {\smallskip}
\multicolumn{5}{c}{Q band$^a$}\\
\hline
32.4 & 9.2 & 54.4 & 0.66 & 0.95 \\
34.6 & 8.6 & 50.9 & 0.62 & 0.95 \\
36.9 & 8.1 & 47.8 & 0.60 & 0.95 \\
39.2 & 7.6 & 45.0 & 0.59 & 0.95 \\
41.5 & 7.2 & 42.5 & 0.58 & 0.95 \\
43.8 & 6.8 & 40.2 & 0.56 & 0.95 \\
46.1 & 6.5 & 38.2 & 0.54 & 0.95 \\
48.4 & 6.2 & 36.4 & 0.51 & 0.93 \\
\hline                       
\multicolumn{5}{c}{W band$^b$}\\
\hline
72.0 & 4.2 & 34.2 & 0.83 & 0.95\\
86.0 & 3.5 & 28.6 & 0.81 & 0.95\\
100.0 & 3.0 & 24.0 & 0.80 & 0.95\\
115.0 & 2.6 & 21.4 & 0.78 & 0.94\\
\hline \noalign {\smallskip}
\end{tabular}
\end{center}
\tablefoot{
	\tablefoottext{a}{QUIJOTE and SANCHO data recorded with the Yebes 40\,m telescope (see Sect.\,\ref{observations}). \texttt{https://rt40m.oan.es/rt40m$\_$en.php}.} 
	\tablefoottext{b}{The 3\,mm survey of TMC-1 recorded with the IRAM 30\,m telescope (see Sect.\,\ref{observations}). \texttt{https://publicwiki.iram.es/Iram30mEfficiencies}.}
}
\end{table}

\begin{table*}
\caption{Spectroscopic and observed line parameters for the doubly substituted isotopologues of HCCCN in TMC-1.}
\label{tab_fits_doubleiso}
\centering
\begin{tabular}{llccl|ccccc|c}
\hline
\multicolumn{1}{c}{Molecule} & \multicolumn{1}{c}{$J_{\rm u}-J_{\rm l}$} & \multicolumn{1}{c}{$\nu_{\rm rest}^a$} & 
\multicolumn{1}{c}{$E_{\rm u}/k_{\rm B}$} &  \multicolumn{1}{c}{$S_{\rm ij}$} & 
\multicolumn{1}{c}{$T_{\rm A}^*$} & \multicolumn{1}{c}{$v_{\rm LSR}^b$}   &  \multicolumn{1}{c}{$\Delta$\,$v^c$} &
\multicolumn{1}{c}{$\int$$T_{\rm A}^* dv$$^d$} & \multicolumn{1}{c}{$\sigma^e$}        & \multicolumn{1}{c}{$T_{\rm ex}^f$}\\
         &                      &           \multicolumn{1}{c}{(MHz)} &   \multicolumn{1}{c}{(K)}&  & \multicolumn{1}{c}{(mK)}     & 
         \multicolumn{1}{c}{(km\,s$^{-1}$)}  & \multicolumn{1}{c}{(km\,s$^{-1}$)} &  \multicolumn{1}{c}{(mK\,km\,s$^{-1}$)}   &
         \multicolumn{1}{c}{(mK)} & \multicolumn{1}{c}{(K)}\\
\hline         
D$^{13}$CCCN       &4$-$3$^g$    & 32857.4694(09) & 3.9 & 2.86     & 0.70 & 5.74(07) & 0.75(17) & 0.56(06) & 0.09 & 7.4\\
D$^{13}$CCCN       &4$-$3$^h$    & 32857.5804(09) & 3.9 & 8.64$^i$ & 2.38 & 5.77(02) & 0.75(05) & 1.90(19) & 0.09 & 7.6/8.8\\
D$^{13}$CCCN       &5$-$4$^j$    & 41071.8709(10) & 5.9 & 14.6$^i$ & 3.67 & 5.78(01) & 0.84(03) & 3.29(33) & 0.12 & 6.4/6.4/6.2\\
D$^{13}$CCCN       &6$-$5$^j$    & 49286.1351(13) & 8.3 & 17.7$^i$ & 3.08 & 5.84(02) & 0.53(05) & 1.75(17) & 0.32 & 5.6/5.6/5.3\\
DC$^{13}$CCN       &4$-$3$^g$    & 33660.2341(06) & 4.0 & 2.86     & 0.94 & 5.86(04) & 0.63(11) & 0.63(06) & 0.08 & 7.3\\
DC$^{13}$CCN       &4$-$3$^h$    & 33660.3452(06) & 4.0 & 8.64$^i$ & 1.87 & 5.78(02) & 0.85(05) & 1.67(17) & 0.08 & 7.4/8.5\\
DC$^{13}$CCN       &5$-$4$^{j,k}$& 42075.3228(08) & 6.1 & 14.6$^i$ & 2.79 & 5.81(02) & 0.82(02) & 2.44(24) & 0.13 & 6.2/6.2/6.0\\
DCC$^{13}$CN       &4$-$3$^{j,l}$& 33620.0314(04) & 4.0 & 11.5$^i$ & ...  & ...      & ...      & ...      & ...  & 7.3/7.4/8.5\\
DCC$^{13}$CN       &5$-$4$^j$    & 42024.9652(06) & 6.1 & 14.6$^i$ & 3.93 & 5.85(01) & 0.71(02) & 2.98(30) & 0.13 & 6.2/6.2/6.0\\
DCCC$^{15}$N       &4$-$3        & 32802.9527(12) & 3.9 & 4.00     & 1.08 & 5.86(03) & 0.61(06) & 0.70(07) & 0.10 & 8.0\\
DCCC$^{15}$N       &5$-$4        & 41003.6144(15) & 5.9 & 5.00     & 1.51 & 5.91(02) & 0.71(06) & 1.14(11) & 0.12 & 6.3\\
DCCC$^{15}$N       &6$-$5        & 49204.2250(18) & 8.3 & 6.00     & 1.19 & 5.88(08) & 0.69(21) & 0.88(09) & 0.33 & 5.5\\
H$^{13}$C$^{13}$CCN&4$-$3$^g$    & 35137.0295(28) & 4.2 & 2.86     & 0.39 & 5.80(16) & 0.58(29) & 0.24(02) & 0.12 & 7.0\\
H$^{13}$C$^{13}$CCN&4$-$3$^h$    & 35137.1406(28) & 4.2 & 8.64$^i$ & 1.60 & 5.73(03) & 0.90(10) & 1.53(15) & 0.12 & 7.1/7.9\\
H$^{13}$C$^{13}$CCN&5$-$4$^j$    & 43921.3068(34) & 6.3 & 14.6$^i$ & 1.97 & 5.74(03) & 0.90(07) & 1.90(19) & 0.17 & 6.0/5.9/5.8\\
H$^{13}$CC$^{13}$CN&4$-$3$^g$    & 35110.0747(05) & 4.2 & 2.86     & 0.50 & 5.57(32) & 1.13(58) & 0.61(06) & 0.10 & 7.0 \\
H$^{13}$CC$^{13}$CN&4$-$3$^h$    & 35110.1857(05) & 4.2 & 8.64$^i$ & 1.73 & 5.70(04) & 0.72(08) & 1.32(13) & 0.10 & 7.1/7.9\\
H$^{13}$CC$^{13}$CN&5$-$4$^j$    & 43887.6131(06) & 6.3 & 14.6$^i$ & 2.54 & 5.78(02) & 0.98(06) & 2.66(26) & 0.18 & 6.0/5.9/5.8\\
HC$^{13}$C$^{13}$CN&4$-$3$^g$    & 36082.3614(07) & 4.3 & 2.86     & 0.74 & 5.70(07) & 0.60(10) & 0.47(05) & 0.10 & 6.8\\
HC$^{13}$C$^{13}$CN&4$-$3$^h$    & 36082.4724(07) & 4.3 & 8.64$^i$ & 2.23 & 5.76(04) & 0.70(08) & 1.66(17) & 0.10 & 6.8/7.6\\
HC$^{13}$C$^{13}$CN&5$-$4$^j$    & 45102.9659(08) & 6.5 & 14.6$^i$ & 2.90 & 5.76(02) & 0.75(02) & 2.32(23) & 0.18 & 5.8/5.8/5.6\\
HCC$^{13}$C$^{15}$N&4$-$3        & 35205.8007(04) & 4.2 & 4.00     & 1.35 & 5.76(03) & 0.79(07) & 1.13(11) & 0.10 & 7.3\\
HCC$^{13}$C$^{15}$N&5$-$4        & 44007.1592(04) & 6.3 & 5.00     & 1.28 & 5.65(05) & 0.78(14) & 1.07(11) & 0.21 & 5.9\\
HC$^{13}$CC$^{15}$N&4$-$3        & 35173.1533(10) & 4.2 & 4.00     & 0.99 & 5.84(04) & 0.73(09) & 0.76(07) & 0.11 & 7.3\\
HC$^{13}$CC$^{15}$N&5$-$4$^m$    & 43966.3500(10) & 6.3 & 5.00     & 1.36 & 5.89(02) & 0.82(06) & 1.19(12) & 0.14 & 5.9\\
\hline
\end{tabular}
\tablefoot{Values in parentheses are the uncertainty of the last significant figures.
	\tablefoottext{a}{Adopted rest frequencies.} 
	\tablefoottext{b}{Local standard of rest velocity of the emission for the adopted rest frequency (in km\,s$^{-1}$).}
	\tablefoottext{c}{Line width at half-intensity derived by fitting a Gaussian profile to the observed
     lines (in km\,s$^{-1}$).}
	\tablefoottext{d}{Integrated line intensity in mK\,km\,s$^{-1}$. We assumed that the uncertainty is dominated by the calibration error of 10\%.}
	\tablefoottext{e}{The sensitivity of the data (root mean square error) has been derived from a baseline fit to each line in a velocity window from $-$10 to 20 km\,s$^{-1}$ (in mK).}
	\tablefoottext{f}{Excitation temperature derived from the LVG modelling for each hyperfine transition.}
	\tablefoottext{g}{$F_{\rm u}-F_{\rm l}=3-2$ component.}
    \tablefoottext{h}{Weighted average of the $F_{\rm u}-F_{\rm l}=4-3$ and $F_{\rm u}-F_{\rm l}=5-4$ hyperfine components.}
    \tablefoottext{i}{Sum of the line strengths of the hyperfine components that contribute to the line.}
    \tablefoottext{j}{Weighted average of the three strongest hyperfine components.}
    \tablefoottext{k}{Partially blended with an unidentified line.}
    \tablefoottext{l}{Fully blended with C$^{13}$CS.}
    \tablefoottext{m}{Partially blended with HCCC$^{13}$CCCCN. The fit to the blended line was made as the sum of two independent lines with free parameters. One of the lines appears with the radial velocity of TMC-1, and we therefore assumed the parameters given for the fit of this line.}
}
\end{table*}

\begin{table*}
\begin{center}
\caption{Observed line parameters for HCCCN in TMC-1.}
\label{tab_fits_normal}
\resizebox{1\textwidth}{!}{
\begin{tabular}{cccrrrl|cllcc|c}
\hline
\multicolumn{1}{c}{Molecule} & \multicolumn{1}{c}{$J_{\rm u}-J_{\rm l}$} & \multicolumn{1}{c}{$F_{\rm u}-F_{\rm l}$} &  &
\multicolumn{1}{c}{$\nu_{\rm rest}^a$} &
\multicolumn{1}{c}{$E_{\rm u}/k_{\rm B}$} &  \multicolumn{1}{c}{$S_{\rm ij}$} &
\multicolumn{1}{c}{$T_{\rm A}^*$} & \multicolumn{1}{c}{$v_{\rm LSR}^b$}   &  \multicolumn{1}{c}{$\Delta$\,$v^c$} &
\multicolumn{1}{c}{$\int$$T_{\rm A}^* dv$$^d$} & \multicolumn{1}{c}{$\sigma^e$} & \multicolumn{1}{c}{$T_{\rm ex}^f$}\\
         &                      &              &        &  \multicolumn{1}{c}{(MHz)}        &  \multicolumn{1}{c}{(K)}&  & \multicolumn{1}{c}{(K)}     &
         \multicolumn{1}{c}{(km\,s$^{-1}$)}  & \multicolumn{1}{c}{(km\,s$^{-1}$)} &  \multicolumn{1}{c}{(K\,km\,s$^{-1}$)}  & \multicolumn{1}{c}{(mK)} &
         \multicolumn{1}{c}{(K)}\\
\hline         
HCCCN       &4-3   & 4-4              &      & 36390.8876(05)      &  4.4 & 0.250     & 0.4410 &  5.79(01) & 0.69(01) & 0.3233(0323) & 0.1 & 8.1         \\   %     0.10\\
HCCCN       &4-3   & 3-2              &      & 36392.2345(03)      &  4.4 & 2.857     & 1.9243 &  5.77(01) & 0.75(01) & 1.5322(1532) & 0.1 & 8.0         \\   %     0.10\\
HCCCN       &4-3   & 4-3/5-4          & $^g$ & 36392.3455(03)      &  4.4 & 8.639$^h$ & 2.5952 &  5.76(01) & 0.82(01) & 2.5952(2595) & 0.1 & 8.6/9.7     \\   %     0.10\\
HCCCN       &4-3   & 3-4              &      & 36392.7382(03)      &  4.4 & 0.004     & 0.0076 &  5.80(01) & 0.71(01) & 0.0058(0006) & 0.1 & 7.4         \\   %     0.10\\
HCCCN       &4-3   & 3-3              &      & 36394.1777(06)      &  4.4 & 0.250     & 0.4404 &  5.78(01) & 0.69(01) & 0.3247(0325) & 0.1 & 7.8         \\   %     0.10\\
HCCCN       &5-4   & 5-5              &      & 45488.8387(05)      &  6.5 & 0.200     & 0.3722 &  5.78(01) & 0.59(01) & 0.2338(0234) & 0.2 & 7.4         \\   %     0.21\\
HCCCN       &5-4   & 4-3/5-4/6-5      & $^g$ & 45490.3063(05)      &  6.5 & 14.60$^h$ & 3.1585 &  5.80(01) & 0.94(01) & 3.1295(3129) & 0.2 & 8.2/8.4/8.8 \\   %     0.21\\
HCCCN       &5-4   & 4-5              &      & 45490.6380(04)      &  6.5 & 0.002     & 0.0038 &  5.81(02) & 0.50(03) & 0.0020(0002) & 0.2 & 6.7         \\   %     0.21\\
HCCCN       &5-4   & 4-4              &      & 45492.1101(06)      &  6.5 & 0.200     & 0.3712 &  5.80(01) & 0.60(01) & 0.2351(0235) & 0.2 & 7.6         \\   %     0.21\\
HCCCN       &8-7   & 8-8              &      & 72782.2941(07)      & 15.7 & 0.125     & 0.2634 &  5.80(01) & 0.50(01) & 0.1413(0141) & 4.1 & 7.4         \\   %     4.13\\
HCCCN       &8-7   & 7-6/8-7/9-8      & $^g$ & 72783.8171(06)      & 15.7 & 23.75$^h$ & 4.5540 &  5.78(01) & 0.60(01) & 2.9206(2921) & 4.1 & 7.9/8.0/8.2 \\   %     4.13\\
HCCCN       &8-7   & 7-7              &      & 72785.5455(08)      & 15.7 & 0.125     & 0.1990 &  5.79(01) & 0.52(01) & 0.1102(0110) & 4.1 & 7.4         \\   %     4.13\\
HCCCN       &9-8   & 9-9              &      & 81879.9284(08)      & 19.6 & 0.111     & 0.1735 &  5.81(01) & 0.47(01) & 0.0866(0087) & 2.9 & 6.6         \\   %     2.95\\
HCCCN       &9-8   & 8-7/9-8/10-9     & $^g$ & 81881.4617(07)      & 19.6 & 26.78$^h$ & 4.3344 &  5.78(01) & 0.53(01) & 2.4576(2457) & 2.9 & 7.1/7.2/7.3 \\   %     2.95\\
HCCCN       &9-8   & 8-8              &      & 81883.1771(08)      & 19.6 & 0.111     & 0.1333 &  5.79(01) & 0.45(01) & 0.0643(0064) & 2.9 & 6.7         \\   %     2.95\\
HCCCN       &10-9  & 10-10            &      & 90977.4469(09)      & 24.0 & 0.100     & 0.0890 &  5.83(01) & 0.46(01) & 0.0436(0043) & 2.7 & 5.8         \\   %     2.70\\
HCCCN       &10-9  & 9-8/10-9/11-10   & $^g$ & 90978.9886(08)      & 24.0 & 29.80$^h$ & 3.4147 &  5.81(01) & 0.51(01) & 1.8423(1842) & 2.7 & 6.1/6.3/6.4 \\   %     2.70\\
HCCCN       &10-9  & 9-9              &      & 90980.6938(09)      & 24.0 & 0.100     & 0.0690 &  5.85(01) & 0.42(01) & 0.0312(0031) & 2.7 & 5.8         \\   %     2.70\\
HCCCN       &11-10 & 11-11            &      & 100074.8364(09)     & 28.8 & 0.091     & 0.0331 &  5.85(01) & 0.42(01) & 0.0147(0015) & 1.1 & 5.0         \\   %     1.09\\
HCCCN       &11-10 & 10-9/11-10/12-11 & $^g$ & 100076.3849(08)     & 28.8 & 32.82$^h$ & 2.2851 &  5.81(01) & 0.49(01) & 1.1864(1186) & 1.1 & 5.3/5.4/5.4 \\   %     1.09\\
HCCCN       &11-10 & 10-10            &      & 100078.0818(10)     & 28.8 & 0.091     & 0.0241 &  5.84(01) & 0.42(02) & 0.0107(0011) & 1.1 & 5.0         \\   %     1.09\\
HCCCN       &12-11 & 12-12            &      & 109172.0832(10)     & 34.1 & 0.083     & 0.0103 &  5.84(05) & 0.25(13) & 0.0027(0011) & 4.4 & 4.5         \\   %     4.38\\
HCCCN       &12-11 & 11-10/12-11/13-12& $^g$ & 109173.6375(10)     & 34.1 & 35.83$^h$ & 1.4581 &  5.83(01) & 0.46(01) & 0.7225(0722) & 4.4 & 4.9/4.8/4.8 \\   %     4.38\\
HCCCN       &12-11 & 11-11            &      & 109175.3275(10)     & 34.1 & 0.083     & 0.0120 &  5.81(06) & 0.19(12) & 0.0025(0011) & 4.4 & 4.6         \\   %     4.38\\
\hline
\end{tabular}
}
\end{center}
\tablefoot{Values in parentheses are the uncertainty of the last significant figures.
    \tablefoottext{a}{Adopted rest frequencies.}
    \tablefoottext{b}{Local standard of rest velocity of the emission for the adopted rest frequency (in km\,s$^{-1}$).}
    \tablefoottext{c}{Line width at half-intensity derived by fitting a Gaussian profile to the observed lines (in km\,s$^{-1}$).}
	\tablefoottext{d}{Integrated line intensity in K\,km\,s$^{-1}$; we assumed that the uncertainty is dominated by the calibration error of 10\%.}
	\tablefoottext{e}{The sensitivity of the data (root mean square error) has been derived from a baseline fit to each line in a velocity window from $-$15 to 25 km\,s$^{-1}$ (in mK).}
	\tablefoottext{f}{Excitation temperature derived from the LVG modelling for each hyperfine transition. The density of the model is different for
	the lines at 7\,mm ($n$(H$_2$)\,=\,1$\times$10$^4$\,cm$^{-3}$) and those at 3\,mm ($n$(H$_2$)\,=\,2$\times$10$^4$\,cm$^{-3}$), see Sect.\,\ref{lvg}.}
    \tablefoottext{g}{Weighted average of the involved hyperfine components.}
    \tablefoottext{h}{Sum of the line strengths of the hyperfine components that contribute to the line.}
}
\end{table*}

\begin{table*}
\begin{center}
\caption{Observed line parameters for singly substituted isotopologues of HCCCN in TMC-1.}
\label{tab_fits_single}
\resizebox{1\textwidth}{!}{
\begin{tabular}{lccrrcl|cllcc|c}
%\begin{tabular}{lcclcllcc}
\hline
\multicolumn{1}{c}{Molecule} & \multicolumn{1}{c}{$J_{\rm u}-J_{\rm l}$} & \multicolumn{1}{c}{$F_{\rm u}-F_{\rm l}$} & & \multicolumn{1}{c}{$\nu_{\rm rest}^a$} &
\multicolumn{1}{c}{$E_{\rm u}/k_{\rm B}$} &  \multicolumn{1}{c}{$S_{\rm ij}$} &
\multicolumn{1}{c}{$T_{\rm A}^*$} & \multicolumn{1}{c}{$v_{\rm LSR}^b$}   &  \multicolumn{1}{c}{$\Delta$\,$v^c$} &
\multicolumn{1}{c}{$\int$$T_{\rm A}^* dv$$^d$} & \multicolumn{1}{c}{$\sigma^e$} & \multicolumn{1}{c}{$T_{\rm ex}^f$}\\
         &                      &             &         &  \multicolumn{1}{c}{(MHz)}
& \multicolumn{1}{c}{(K)}& \multicolumn{1}{c}{} & \multicolumn{1}{c}{(K)} &
         \multicolumn{1}{c}{(km\,s$^{-1}$)}  & \multicolumn{1}{c}{(km\,s$^{-1}$)} &  \multicolumn{1}{c}{(K\,km\,s$^{-1}$)}   &
         \multicolumn{1}{c}{(mK)} & \multicolumn{1}{c}{(K)}\\
\hline
DCCCN        &4-3  & 4-4              &      & 33771.0873(03)& 4.1 & 0.25     & 0.00715    &  5.82(1) & 0.75(01) & 0.0057(006) & 0.09 & 6.6\\
DCCCN        &4-3  & 3-2              &      & 33772.4373(03)& 4.1 & 2.86     & 0.08011    &  5.71(1) & 0.91(01) & 0.0775(078) & 0.09 & 7.3\\
DCCCN        &4-3  & 4-3/5-4          & $^g$ & 33772.5500(03)& 4.1 & 8.64$^h$ & 0.21635    &  5.80(1) & 0.81(01) & 0.1861(186) & 0.09 & 7.5/8.5\\
DCCCN        &4-3  & 3-3              &      & 33774.3772(03)& 4.1 & 0.25     & 0.00470    &  5.84(1) & 0.76(02) & 0.0038(004) & 0.09 & 6.8\\
DCCCN        &5-4  & 5-5              &      & 42214.1049(04)& 6.1 & 0.20     & 0.00546    &  5.81(1) & 0.62(02) & 0.0036(004) & 0.12 & 5.4\\
DCCCN        &5-4  & 4-3/5-4/6-5      & $^g$ & 42215.5779(03)& 6.1 & 14.6$^h$ & 0.28307    &  5.80(1) & 0.82(01) & 0.2482(248) & 0.12 & 6.3/6.3/6.2\\
DCCCN        &5-4  & 4-4              &      & 42217.3762(04)& 6.1 & 0.20     & 0.00530    &  5.83(1) & 0.60(01) & 0.0034(003) & 0.12 & 6.0\\
DCCCN        &9-8  & 8-7/9-8/10-9     & $^g$ & 75987.1367(06)&18.2 & 26.8$^h$ & 0.18741    &  5.83(1) & 0.53(01) & 0.1054(105) & 2.80 & 4.6/4.6/4.5\\
DCCCN        &10-9 & 9-8/10-9/11-10   & $^g$ & 84429.8088(07)&22.3 & 29.8$^h$ & 0.09951    &  5.83(1) & 0.53(01) & 0.0562(056) & 0.81 & 4.5/4.5/4.5\\
DCCCN        &11-10& 10-9/11-10/12-11 & $^g$ & 92872.3725(07)&26.7 & 32.8$^h$ & 0.04898    &  5.85(1) & 0.49(01) & 0.0253(025) & 0.92 & 4.6/4.6/4.6\\
DCCCN        &12-11& 11-10/12-11/13-12& $^g$ &101314.8168(08)&31.6 & 35.8$^h$ & 0.01870    &  5.84(2) & 0.42(04) & 0.0084(008) & 2.04 & 4.6/4.6/4.6\\
DCCCN        &13-12& 12-11/13-12/14-13& $^g$ &109757.1310(08)&36.8 & 38.8$^h$ & 0.02084    &  5.82(4) & 0.37(08) & 0.0082(016) & 4.89 & 4.7/4.7/4.7\\
\\
H$^{13}$CCCN &4-3  & 4-4              &      & 35265.9791(49)& 4.2 & 0.25     & 0.00437    &  5.95(2) & 0.62(04) & 0.0031(003) & 0.14 & 6.2 \\
H$^{13}$CCCN &4-3  & 3-2              &      & 35267.3117(07)& 4.2 & 2.86     & 0.05278    &  5.79(1) & 0.72(01) & 0.0402(040) & 0.14 & 7.0 \\
H$^{13}$CCCN &4-3  & 4-3/5-4          & $^g$ & 35267.4215(06)& 4.2 & 8.64$^h$ & 0.14143    &  5.77(1) & 0.76(01) & 0.1148(115) & 0.14 & 7.1/7.9\\
H$^{13}$CCCN &4-3  & 3-3              &      & 35269.2342(62)& 4.2 & 0.25     & 0.00482    &  5.63(1) & 0.71(02) & 0.0036(004) & 0.14 & 6.5\\
H$^{13}$CCCN &5-4  & 5-5              &      & 44082.7057(50)& 6.3 & 0.20     & 0.00367    &  5.91(2) & 0.65(04) & 0.0026(003) & 0.21 & 5.1\\
H$^{13}$CCCN &5-4  & 4-3/5-4/6-5      & $^g$ & 44084.1577(08)& 6.3 & 14.6$^h$ & 0.17323    &  5.78(1) & 0.79(01) & 0.1456(146) & 0.21 & 6.0/6.0/5.8\\
H$^{13}$CCCN &5-4  & 4-4              &      & 44085.9423(61)& 6.3 & 0.20     & 0.00306    &  5.69(2) & 0.55(04) & 0.0018(002) & 0.21 & 5.6\\
H$^{13}$CCCN &9-8  & 8-7/9-8/10-9     & $^g$ & 79350.4624(13)&19.0 & 26.8$^h$ & 0.08837    &  5.81(1) & 0.50(01) & 0.0469(047) & 2.01 & 4.5/4.5/4.4\\
H$^{13}$CCCN &10-9 & 9-8/10-9/11-10   & $^g$ & 88166.7930(14)&23.3 & 29.8$^h$ & 0.04218    &  5.80(1) & 0.44(02) & 0.0195(020) & 2.55 & 4.5/4.5/4.5\\
H$^{13}$CCCN &11-10& 10-9/11-10/12-11 & $^g$ & 96983.0010(16)&27.9 & 32.8$^h$ & 0.01794    &  5.82(1) & 0.54(02) & 0.0103(010) & 0.64 & 4.6/4.6/4.5\\
H$^{13}$CCCN &12-11& 11-10/12-11/13-12& $^g$ &105799.0742(17)&33.0 & 35.8$^h$ & 0.00661    &  5.59(7) & 0.26(13) & 0.0018(009) & 3.31 & 4.6/4.6/4.6\\
H$^{13}$CCCN &13-12& 12-11/13-12/14-13& $^g$ &114615.0008(19)&38.5 & 38.8$^h$ & $^i$       &  ...     & ...      & ...           & 7.04 & 4.7/4.7/4.7\\
\\
HC$^{13}$CCN &4-3  & 4-4              &      & 36236.5139(53)& 4.3 & 0.25     & 0.00508    &  5.87(1) & 0.71(02) & 0.0038(004) & 0.11 & 6.1 \\
HC$^{13}$CCN &4-3  & 3-2              &      & 36237.8536(07)& 4.3 & 2.86     & 0.05919    &  5.76(1) & 0.74(01) & 0.0467(047) & 0.11 & 6.8 \\
HC$^{13}$CCN &4-3  & 4-3/5-4          & $^g$ & 36237.9641(07)& 4.3 & 8.64$^h$ & 0.15201    &  5.74(1) & 0.70(01) & 0.1132(113) & 0.11 & 6.8/7.6\\
HC$^{13}$CCN &4-3  & 3-3              &      & 36239.7865(68)& 4.3 & 0.25     & 0.00499    &  5.72(1) & 0.67(02) & 0.0036(004) & 0.11 & 6.3\\
HC$^{13}$CCN &5-4  & 5-5              &      & 45295.8703(54)& 6.5 & 0.20     & 0.00326    &  5.87(2) & 0.53(04) & 0.0018(002) & 0.17 & 5.0\\
HC$^{13}$CCN &5-4  & 4-3/5-4/6-5      & $^g$ & 45297.3301(08)& 6.5 & 14.6$^h$ & 0.18974    &  5.78(1) & 0.78(01) & 0.1573(157) & 0.17 & 5.8/5.8/5.7\\
HC$^{13}$CCN &5-4  & 4-4              &      & 45299.1243(67)& 6.5 & 0.20     & 0.00370    &  5.73(1) & 0.55(02) & 0.0022(002) & 0.17 & 5.5\\
HC$^{13}$CCN &8-7  & 7-6/8-7/9-8      & $^g$ & 72475.0586(13)&15.7 & 23.7$^h$ & 0.16345$^j$&  5.80(2) & 0.53(03) & 0.0924(092) & 3.51 & 4.5/4.5/4.5\\
HC$^{13}$CCN &9-8  & 8-7/9-8/10-9     & $^g$ & 81534.1100(14)&19.6 & 26.8$^h$ & 0.07847    &  5.82(1) & 0.56(01) & 0.0465(046) & 0.77 & 4.5/4.4/4.4\\
HC$^{13}$CCN &10-9 & 9-8/10-9/11-10   & $^g$ & 90593.0444(15)&23.9 & 29.8$^h$ & 0.04196    &  5.83(1) & 0.50(02) & 0.0222(022) & 1.73 & 4.5/4.5/4.4\\
HC$^{13}$CCN &11-10& 10-9/11-10/12-11 & $^g$ & 99651.8487(17)&28.7 & 32.8$^h$ & 0.01884    &  5.82(1) & 0.59(02) & 0.0191(019) & 0.95 & 4.6/4.5/4.5\\
HC$^{13}$CCN &12-11& 11-10/12-11/13-12& $^g$ &108710.5101(18)&33.9 & 35.8$^h$ & 0.01088    &  5.89(8) & 0.34(17) & 0.0040(013) & 3.70 & 4.6/4.6/4.6\\
\\
HCC$^{13}$CN &4-3  & 4-4              &      & 36240.0103(24)& 4.3 & 0.25     & 0.00766    &  5.83(1) & 0.70(01) & 0.0057(006) & 0.12 & 6.1\\
HCC$^{13}$CN &4-3  & 3-2              &      & 36241.3514(07)& 4.3 & 2.86     & 0.08741    &  5.74(1) & 0.74(01) & 0.0689(069) & 0.12 & 6.8\\
HCC$^{13}$CN &4-3  & 4-3/5-4          & $^g$ & 36241.4620(07)& 4.3 & 8.64$^h$ & 0.22545    &  5.78(1) & 0.79(01) & 0.1899(190) & 0.12 & 6.9/7.6\\
HCC$^{13}$CN &4-3  & 3-3              &      & 36243.2863(30)& 4.3 & 0.25     & 0.00791    &  5.72(1) & 0.70(01) & 0.0060(006) & 0.12 & 6.3\\
HCC$^{13}$CN &5-4  & 5-5              &      & 45300.2411(25)& 6.5 & 0.20     & 0.00547    &  5.84(1) & 0.62(02) & 0.0036(004) & 0.15 & 5.0\\
HCC$^{13}$CN &5-4  & 4-3/5-4/6-5      & $^g$ & 45301.7024(09)& 6.5 & 14.6$^h$ & 0.27014    &  5.76(1) & 0.78(01) & 0.2249(225) & 0.15 & 5.8/5.8/5.7\\
HCC$^{13}$CN &5-4  & 4-4              &      & 45303.4985(30)& 6.5 & 0.20     & 0.00534    &  5.74(1) & 0.60(03) & 0.0034(003) & 0.15 & 5.5\\
HCC$^{13}$CN &8-7  & 7-6/8-7/9-8      & $^g$ & 72482.0540(14)&15.7 & 23.7$^h$ & 0.22458$^j$&  5.77(2) & 0.57(03) & 0.1368(137) & 4.22 & 4.5/4.5/4.5\\
HCC$^{13}$CN &9-8  & 8-7/9-8/10-9     & $^g$ & 81541.9798(15)&19.6 & 26.8$^h$ & 0.11192    &  5.80(1) & 0.55(01) & 0.0654(065) & 0.82 & 4.5/4.4/4.4\\
HCC$^{13}$CN &10-9 & 9-8/10-9/11-10   & $^g$ & 90601.7884(16)&23.9 & 29.8$^h$ & 0.06330    &  5.80(1) & 0.49(01) & 0.0328(033) & 1.54 & 4.5/4.5/4.4\\
HCC$^{13}$CN &11-10& 10-9/11-10/12-11 & $^g$ & 99661.4670(18)&28.7 & 32.8$^h$ & 0.02553    &  5.77(1) & 0.50(02) & 0.0135(014) & 0.93 & 4.6/4.5/4.5\\
HCC$^{13}$CN &12-11& 11-10/12-11/13-12& $^g$ &108721.0025(19)&33.9 & 35.8$^h$ & 0.01105$^j$&  5.80(7) & 0.49(17) & 0.0058(015) & 3.61 & 4.6/4.6/4.6\\
\\           
HCCC$^{15}$N &4-3  &                  &      & 35333.8879(07)& 4.2 & 4.00     & 0.07884    &  5.76(1) & 0.68(01) & 0.0573(057) & 0.12 & 7.3 \\
HCCC$^{15}$N &5-4  &                  &      & 44167.2678(09)& 6.4 & 5.00     & 0.08333    &  5.77(1) & 0.60(01) & 0.0532(053) & 0.16 & 5.9 \\
HCCC$^{15}$N &9-8  &                  &      & 79500.0510(15)&19.1 & 9.00     & 0.02823    &  5.75(2) & 0.51(04) & 0.0154(015) & 2.05 & 4.5 \\
HCCC$^{15}$N &10-9 &                  &      & 88333.0013(17)&23.3 & 10.0     & 0.01511    &  5.75(2) & 0.39(06) & 0.0063(008) & 2.12 & 4.5\\
HCCC$^{15}$N &11-10&                  &      & 97165.8288(19)&28.0 & 11.0     & 0.00537    &  5.74(4) & 0.39(08) & 0.0022(004) & 1.11 & 4.6 \\
HCCC$^{15}$N &12-11&                  &      &105998.5213(20)&33.1 & 12.0     & $^i$       &  ...     & ...      & ...         & 3.59 & 4.6 \\
HCCC$^{15}$N &13-12&                  &      &114831.0666(22)&38.6 & 13.0     & $^i$       &  ...     & ...      & ...         & 7.22 & 4.9 \\
\hline
\end{tabular}
}
\end{center}
\tablefoot{Values in parentheses are the uncertainty of the last significant figures.
    \tablefoottext{a}{Adopted rest frequencies.}
    \tablefoottext{b}{Local standard of rest velocity of the emission for the adopted rest frequency (in km\,s$^{-1}$).} 
    \tablefoottext{c}{Line width at half-intensity derived by fitting a Gaussian profile to the observed lines (in km\,s$^{-1}$).}
	\tablefoottext{d}{Integrated line intensity in mK\,km\,s$^{-1}$. We assumed that the uncertainty is dominated by the calibration error of 10\%.}
	\tablefoottext{e}{The sensitivity of the data (root mean square error) has been derived from a baseline fit to each line in a velocity window from $-$15 to 25 km\,s$^{-1}$ (in mK).}
	\tablefoottext{f}{Excitation temperature derived from the LVG modelling for each hyperfine transition.}
    \tablefoottext{g}{Weighted average of the involved hyperfine components.}
    \tablefoottext{h}{Sum of the line strengths of the hyperfine components that contribute to the line.}
    \tablefoottext{i}{Below the detection limit.}
    \tablefoottext{j}{Partially blended with a negative feature produced in the data folding of the frequency-switching observation procedure.}
}
\end{table*}

\begin{table*}
%\begin{center}
\caption{Column densities ($N$) and derived isotopic ratios for the HC$_3$N isotopologues towards TMC-1 (CP).}
\label{table_cd}
\resizebox{1\textwidth}{!}{
\centering
\begin{tabular}{lclccccccc}   
\hline
Species & $N$ (cm$^{-2}$) & \multicolumn{1}{c}{$N$ (cm$^{-2}$)} & \multicolumn{6}{c}{Isotopic ratios$^a$} \\
 & (LVG) & \multicolumn{1}{c}{(LTE)} & D/H & \multicolumn{1}{c}{C/$^{13}$C} & N/$^{15}$N & D/$^{13}$C & D/$^{15}$N & $^{13}$C/$^{15}$N \\
\hline
\hline

HCCCN               &  $1.90(03)\times10^{14}$ & $3.62(75)\times10^{14}$   & ...           & \multicolumn{1}{c}{...} & ... & ... & ... & ... \\
DCCCN               &  $3.30(04)\times10^{12}$ & $5.93(29)\times10^{12}$   & 0.0174(3)$^b$ & \multicolumn{1}{c}{...} & ... & ... & ... & ...\\
H$^{13}$CCCN        &  $1.80(06)\times10^{12}$ & $5.15(22)\times10^{12}$  & ...           & 106(4)$^b$            & ...         & 1.83(6)$^g$             & ...         & 3.000(7)$^f$ \\
HC$^{13}$CCN        &  $2.00(04)\times10^{12}$ & $3.75(17)\times10^{12}$  & ...           & 95(2)$^b$             & ...         & 1.65(4)$^g$             & ...         & 3.333(7)$^f$ \\
HCC$^{13}$CN        &  $2.80(04)\times10^{12}$ & $6.46(08)\times10^{12}$  & ...           & 69(1)$^b$             & ...         & 1.18(2)$^g$             & ...         & 4.667(8)$^f$ \\
HCCC$^{15}$N        &  $0.60(01)\times10^{12}$ & $1.53(03)\times10^{12}$  & ...           & ...                   & 317(4)$^b$  & ...                     & 5.50(7)$^g$ & ...\\
D$^{13}$CCCN        &  $3.20(12)\times10^{10}$ & \multicolumn{1}{c}{...} & 0.0178(9)$^c$ & 103(4)$^g$            & ...         & 1.7(2)$^h$; 1.4(1)$^i$  & ...         & 3.2(3)$^p$ \\
DC$^{13}$CCN        &  $2.50(11)\times10^{10}$ & \multicolumn{1}{c}{...} & 0.0125(6)$^d$ & 132(6)$^g$            & ...         & 1.3(1)$^h$; 1.00(8)$^j$ & ...         & 2.5(2)$^p$ \\
DCC$^{13}$CN        &  $3.50(12)\times10^{10}$ & \multicolumn{1}{c}{...} & 0.0125(5)$^e$ & 94(3)$^g$             & ...         & 1.5(1)$^i$; 1.4(1)$^j$  & ...         & 3.5(3)$^p$ \\
DCCC$^{15}$N        &  $1.00(09)\times10^{10}$ & \multicolumn{1}{c}{...} & 0.017(2)$^f$  & ...                   & 330(31)$^g$ & 0.9(2)$^k$; 1.2(2)$^l$  & ...         & ... \\
H$^{13}$C$^{13}$CCN &  $1.90(16)\times10^{10}$ & \multicolumn{1}{c}{...} & ...           & 95(9)$^c$; 105(9)$^d$ & ...         & 1.7(2)$^m$; 1.3(1)$^n$  & ...         & 2.4(3)$^l$ \\
H$^{13}$CC$^{13}$CN &  $2.30(16)\times10^{10}$ & \multicolumn{1}{c}{...} & ...           & 78(6)$^c$; 123(9)$^e$ & ...         & 1.4(1)$^m$; 1.5(1)$^o$  & ...         & 2.1(4)$^k$  \\
HC$^{13}$C$^{13}$CN &  $2.50(16)\times10^{10}$ & \multicolumn{1}{c}{...} & ...           & 80(5)$^d$; 112(7)$^e$ & ...         & 1.08(8)$^n$; 1.4(1)$^o$ & ...         & 2.3(4)$^k$; 3.1(4)$^l$ \\
HCC$^{13}$C$^{15}$N &  $1.10(18)\times10^{10}$ & \multicolumn{1}{c}{...} & ...           & 55(9)$^f$             & 255(42)$^e$ & 0.9(2)$^p$              & 3.2(5)$^o$ & ... \\
HC$^{13}$CC$^{15}$N &  $0.80(09)\times10^{10}$ & \multicolumn{1}{c}{...} & ...           & 75(8)$^f$             & 250(28)$^d$ & 1.3(2)$^p$              & 3.1(4)$^n$ & ... \\
\hline
\end{tabular}
}
\tablefoot{Values in parentheses are the uncertainty of the last significant figures.
    \tablefoottext{a}{Column density ratios calculated using the derived LVG column densities.} 
    \tablefoottext{b}{Using HCCCN.} 
    \tablefoottext{c}{Using H$^{13}$CCCN.}
    \tablefoottext{d}{Using HC$^{13}$CCN.}
    \tablefoottext{e}{Using HCC$^{13}$CN.}
    \tablefoottext{f}{Using HCCC$^{15}$N.}
    \tablefoottext{g}{Using DCCCN.}
    \tablefoottext{h}{Using H$^{13}$C$^{13}$CCN.}
    \tablefoottext{i}{Using H$^{13}$CC$^{13}$CN.}
    \tablefoottext{j}{Using HC$^{13}$C$^{13}$CN.}
    \tablefoottext{k}{Using HCC$^{13}$C$^{15}$N.}
    \tablefoottext{l}{Using HC$^{13}$CC$^{15}$N.}
    \tablefoottext{m}{Using D$^{13}$CCCN.}
    \tablefoottext{n}{Using DC$^{13}$CCN.}
    \tablefoottext{o}{Using DCC$^{13}$CN.}
    \tablefoottext{p}{Using DCCC$^{15}$N.}
    }
%\end{center}
\end{table*}

%\section{Collisional cross sections}
\section{Rotational diagrams}
\label{lte}

\begin{figure*}%[h!]
%\centering
%\vspace{1cm}
%\vspace{0.3cm}
\includegraphics[scale=0.40, angle=0]{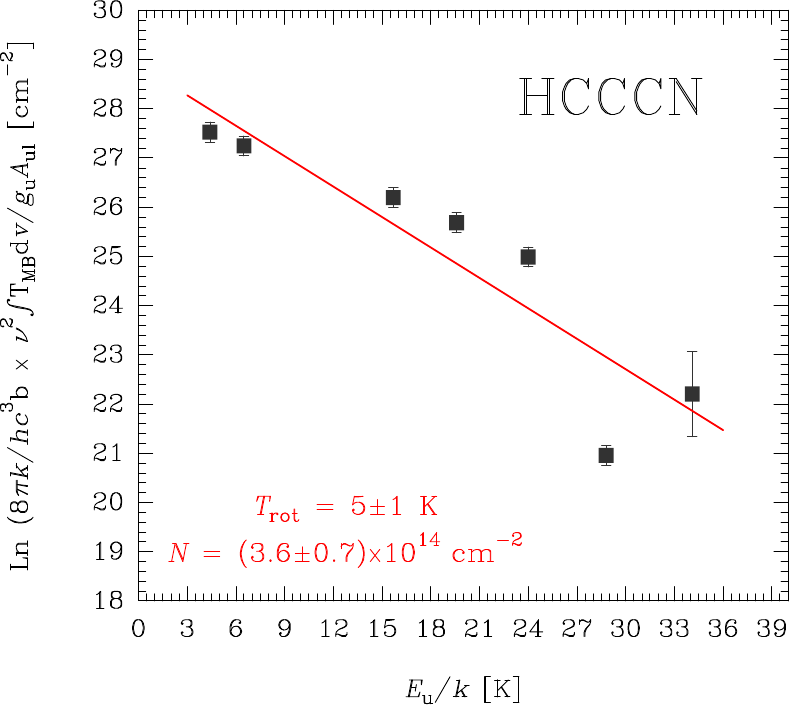}
\hspace{0.5cm}
\includegraphics[scale=0.40, angle=0]{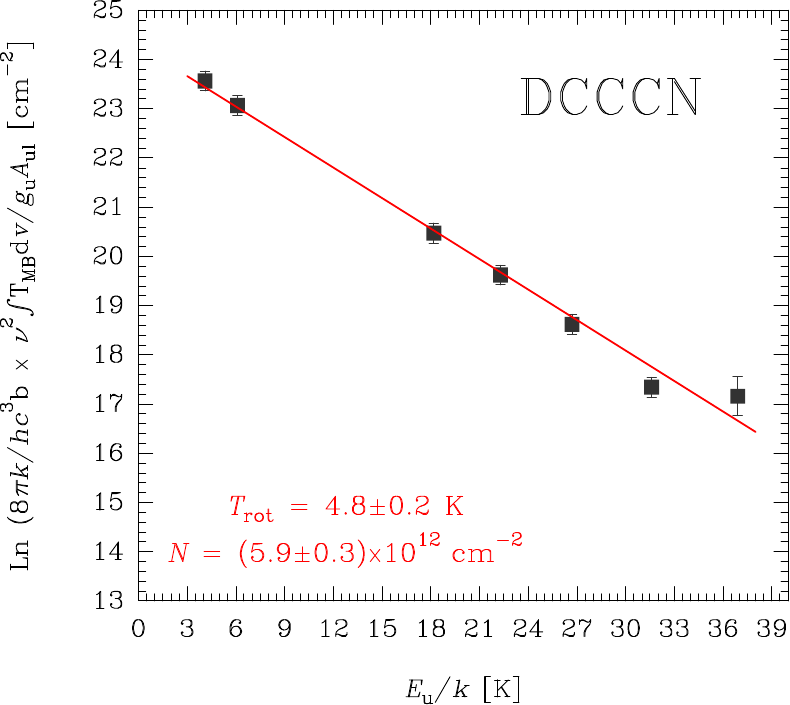} 
\hspace{0.5cm}
\includegraphics[scale=0.40, angle=0]{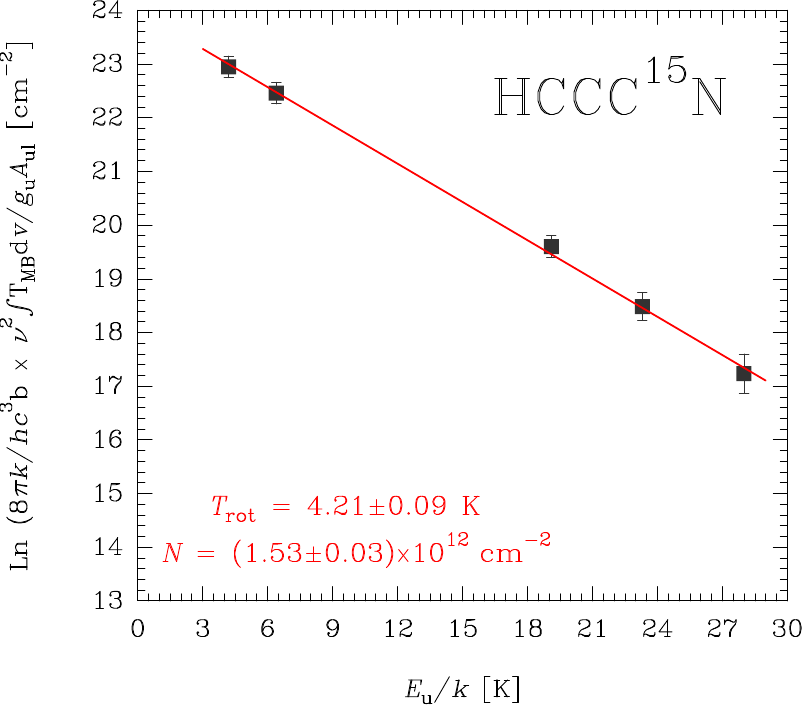} \\
%\vspace{1cm} 
\\
\includegraphics[scale=0.40, angle=0]{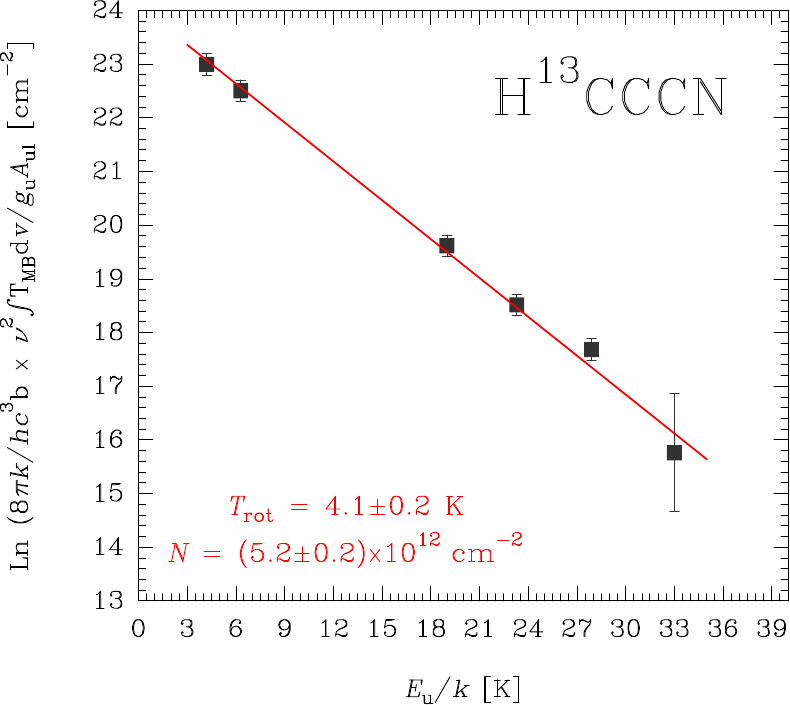} 
\hspace{0.5cm}
\includegraphics[scale=0.40, angle=0]{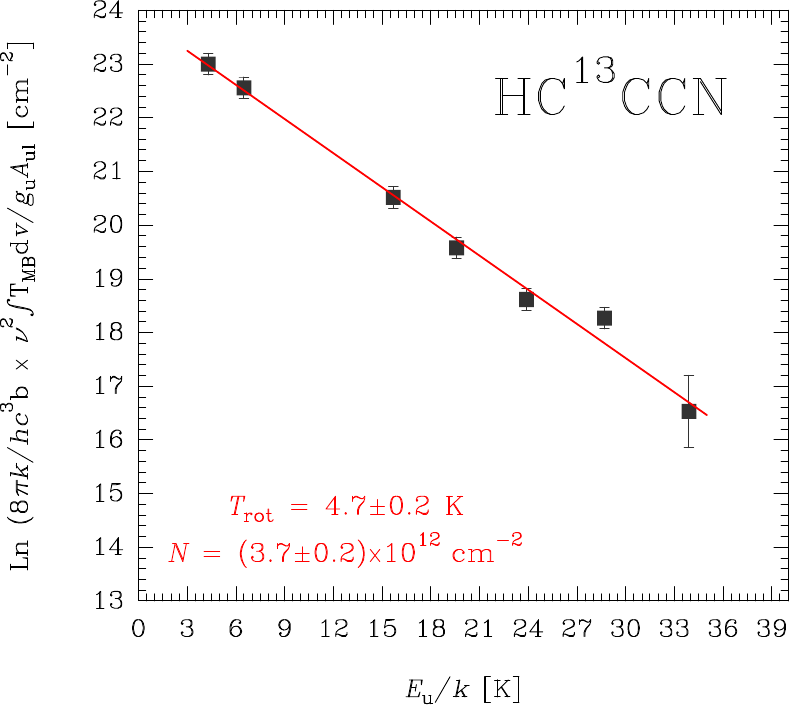}
\hspace{0.5cm}
\includegraphics[scale=0.40, angle=0]{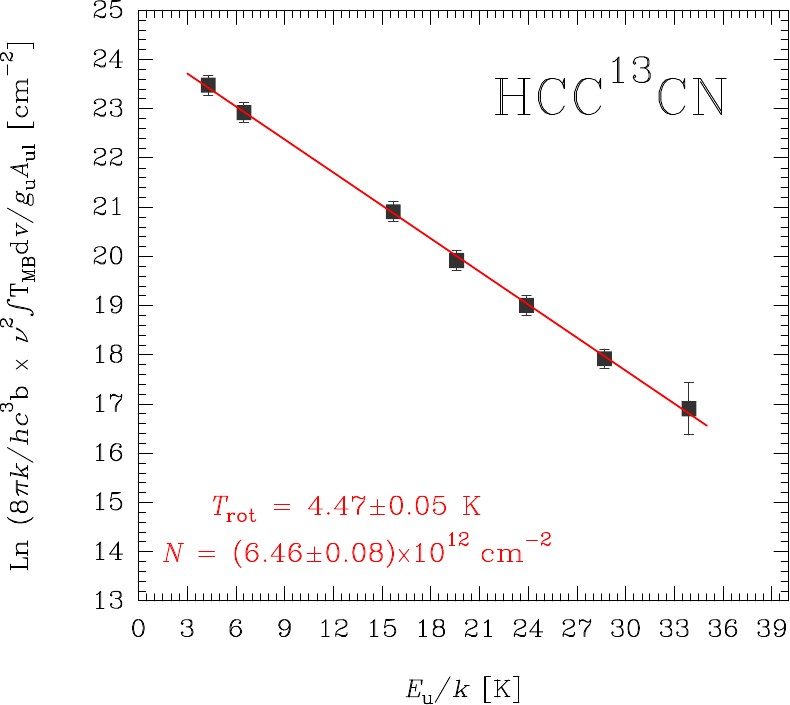}
\caption{Rotational diagrams of HCCCN and its singly substituted isotopologues towards TMC-1 (CP).
The derived values of the rotational temperature, $T_{\rm rot}$, the column density, $N$, and their respective uncertainties are indicated for each species.}\label{fig_DR}
\end{figure*}

To evaluate the rotational temperature ($T_{\rm rot}$), and thus, the level of departure from LTE,
and to obtain an independent estimate of the column density ($N$), we constructed rotation diagrams (see e.g. \citealt{Goldsmith1999})
for all species detected at both 3\,mm and 7\,mm (HCCCN, DCCCN, H$^{13}$CCCN, HC$^{13}$CCN, HCC$^{13}$CN, and HCCC$^{15}$N) for which a large number of lines are detected.
For the double isotopologues, only two lines are detected, so that the rotational diagram method does not provide
a reliable determination of the column density and the rotational temperature.

This analysis assumes the Rayleigh-Jeans approximation, optically thin lines,
and local thermodynamic equilibrium conditions. 
The equation that derives the total column density under these conditions
can be re-arranged as

\begin{equation}
{\rm \ln} \left(\frac{8 \pi k_{\rm B} \nu^2 \int{T_{\rm MB} dv}}{h c^3 A_{\rm ul} g_{\rm u} b}\right) = {\rm \ln} \left(\frac{N}{Q_{\rm rot}} \frac{T_{\rm rot}-T_{\rm bg}}{T_{\rm rot}}\right) - \frac{E_{\rm u}}{k_{\rm B} T_{\rm rot}}
\label{eq_RD}
,\end{equation}
where $g_u$ is the statistical weight in the upper level,
$A_{\rm ul}$ is the Einstein $A$-coefficient for
spontaneous emission, $Q_{\rm rot}$ is the rotational partition
function (which depends on $T_{\rm rot}$), $E_{\rm u}$ is the upper level energy, $\nu$ is the frequency
of the transition, $b$ is the dilution factor, and $T_{\rm bg}$ is the cosmic microwave background radiation temperature.
The first term of Eq.\,(\ref{eq_RD}), which only depends on spectroscopic and
observational line parameters, is plotted as a function of $E_{\rm u}$/$k_{\rm B}$
for the different detected lines. Thus, $T_{\rm rot}$ and $N$ can be derived
by performing a linear least-squares fit to the points (see Fig.\,\ref{fig_DR}).

For HCCCN, DCCCN, H$^{13}$CCCN, HC$^{13}$CCN, and HCC$^{13}$CN different hyperfine structure components of the same
$J_{\rm u}-J_{\rm l}$
transition are observed or are blended in a single line, depending on the frequency of the transition.
Thus, to correctly determine $T_{\rm rot}$ and $N$, 
the integrated intensity ($\mathrm{\int}T_{\rm MB} dv$),
the level degeneracy ($g$), and the line strength ($S$) were calculated as the sum of all 
hyperfine components of each $J_{\rm u}-J_{\rm l}$ transition. The characteristic frequency ($\nu$)
was determined using the weighted average with the relative strength of each hyperfine line as weight,
and the Einstein coefficient ($A$) was calculated using the usual relation.
Due to opacity effects for the different hyperfine components of HCCCN, to derive a reliable
column density for this species, the observed integrated intensity for each $J_{\rm u}-J_{\rm l}$ transition was assumed as the
integrated intensity of the $S_{\rm r}$ satellite line multiplied by a factor 
of the inverse of the relative
intensity (according to the different line strengths) of this hyperfine line with respect to the remaining components of the same $J_{\rm u}-J_{\rm l}$ transition.

The results for $T_{\rm rot}$ and $N$ using the population diagram procedure are shown in 
Table\,\ref{table_cd} and Fig.\,\ref{fig_DR}.
The uncertainties were calculated using the statistical
errors given by the linear least-squares fit for the slope and the
intercept. The individual errors of the data points are
those derived by least squares, taking into account the uncertainty obtained in the determination 
of the observed integrated intensity (see Tables\,\ref{tab_fits_doubleiso}, \ref{tab_fits_normal}, and \ref{tab_fits_single}).

We obtained rotational temperatures 
between 4.1\,K and 5.0\,K for HCCCN and its single isotopologues (see Fig.\,\ref{fig_DR}),
indicating that they are subthermally excited, like most of the species with three to five atoms 
in this region and in cold dark clouds
(see e.g. \citealt{Cernicharo2020a, Cernicharo2020c, Marcelino2020, Agundez2023}), 
and consistent with the derived H$_2$ density. 
We find higher excitation temperatures, closer to the kinetic temperature of 10\,K,
for the $J=4-3$ and $J=5-4$ lines using the LVG formalism (see Tables\,\ref{tab_fits_doubleiso}, \ref{tab_fits_normal}, and \ref{tab_fits_single}).
However, for the lines at 3\,mm, the derived excitation temperatures are about 4-5\,K.
As the excitation temperature derived from the rotational diagrams is determined by the slope, which 
is dominated by the
3\,mm lines, we found these large differences between the excitation temperatures derived with the rotational diagrams and
the kinetic temperature assumed for the LVG modelling.
%The excitation temperatures derived from the LVG model and the rotational temperatures
%from the rotational diagrams agree.
Moreover, in these diagrams, the correction for the background temperature introduces a non-linear dependence
that might introduce some uncertainty in the determination of the rotational temperatures, especially for
low-excitation temperatures.
On the other hand, the column densities derived using this method are dominated by the contribution of the 
low-energy $J=4-3$ and $J=5-4$ lines at 7\,mm.
The column densities derived through the rotation diagram 
are systematically higher by $\sim30-50$\,\% than those derived through the LVG analysis (see Table\,\ref{table_cd}).
These differences mainly arise because various assumptions 
made in the frame of the rotation diagram method break down (see \citealt{Agundez2023}).
We therefore adopted as preferred values
for the column densities the values that were derived through the LVG method, which should also provide more accurate values as long
as the collision rate coefficients with para-H$_2$ and the gas kinetic temperature are known accurately.

\end{appendix}

\end{document}